\documentclass[10pt,aps,prb,twocolumn,superscriptaddress,floatfix]{revtex4-1}

\usepackage{makecell}

\usepackage{amsmath}
\usepackage{amssymb}
\usepackage{graphicx}
\usepackage[free-standing-units=true]{siunitx}

\usepackage{dsfont}

\usepackage{todonotes}
\usepackage[english]{babel}

\usepackage{array}

\usepackage{mathtools}
\usepackage{upgreek}

\usepackage{dsfont}

\newcommand{\sub}[1]{_{\mathrm{#1}}}
\newcommand{\bias}{V\sub{b}}
\newcommand{\Jfreq}{\nu\sub{J}}
\newcommand{\freq}{f}
\newcommand{\reso}{f\sub{0}}
\newcommand{\Icrit}{I\sub{c}}

\newcommand{\PE}{\mathit{P}}

\newcommand{\imp}{\ensuremath{Z\sub{R}(\freq)}}

\newcommand{\CExtractedfig}{\SI{50}{\femto\farad}}
\newcommand{\LpExtracted}{\SI{53}{\pico \henry}}

\newcommand{\TeffExtracted}{\SI{20.9}{\milli\kelvin}}
\newcommand{\TeffFit}{\SI{21}{\milli\kelvin}}
\newcommand{\IsExtracted}{\SI{0.82}{\nano\ampere}}
\newcommand{\IsFit}{\SI{0.85}{\nano\ampere}}
\newcommand{\EcExtracted}{\SI{1.5}{\giga\hertz}}

\newcommand{\CExtracted}{\SI{56.7}{\femto\farad}}

\newcommand{\FreqFit}{\SI{6}{\giga\hertz}}

\newcommand{\RExtracted}{\SI{32.1}{\kilo\ohm}}
\newcommand{\Zzerofit}{\SI{110}{\ohm}}
\newcommand{\Zonefit}{\SI{22}{\ohm}}
\newcommand{\eqcite}[1]{equation~(\ref{#1})}
\newcommand{\figcite}[2]{Fig.~\ref{#1}#2}
\newcommand{\suppcite}[1]{~(Supplementary Information)}
\newcommand{\circuit}{a}
\newcommand{\rc}{c}
\newcommand{\squid}{d}
\newcommand{\chip}{b}
\newcommand{\tunnel}{e}
\newcommand{\notunnel}{f}
\newcommand{\psdfreq}{a}
\newcommand{\impedancepeak}{b}
\newcommand{\psdflux}{c}

\newcommand{\overview}{1}
\newcommand{\psd}{2}
\newcommand{\gfree}{3}
\newcommand{\gpulsed}{4}

\DeclareSIUnit\dBm{dBm}
\DeclareSIUnit\dBc{dBc}
\DeclareSIUnit\photon{photons}
\DeclareSIUnit\sample{S}
\DeclareSIUnit{\sqrthz}{\ensuremath{\sqrt{\text{\hertz}}}}
\DeclareSIUnit{\sqrtsec}{\ensuremath{\sqrt{\text{\second}}}}
\DeclareSIUnit{\sqrtVpercm}{\ensuremath{\sqrt{\text{\volt \per \centi \meter}}}}
\DeclareSIUnit\bit{bit}

\newcommand{\imag}{\imath}

\newenvironment{ac}
{\begin{center}
		\textsc{author contributions}\\[1ex]
	\end{center}
}

\begin{document}
	
	
	\title{A bright on-demand source of anti-bunched microwave photons based on inelastic Cooper pair tunneling}
	
	\author{A. \surname{Grimm}}
	\email{email: alexander.grimm@yale.edu}
	\affiliation{Univ.~Grenoble Alpes, CEA, INAC-PHELIQS, F-38000 Grenoble, France}
	\author{F. \surname{Blanchet}}
	\affiliation{Univ.~Grenoble Alpes, CEA, INAC-PHELIQS, F-38000 Grenoble, France}
	\author{R. \surname{Albert}}
	\affiliation{Univ.~Grenoble Alpes, CEA, INAC-PHELIQS, F-38000 Grenoble, France}
	\author{J. \surname{Lepp\"akangas}}
	\affiliation{Physikalisches Institut, Karlsruhe Institute of Technology, 76131 Karlsruhe, Germany}
	\author{S. \surname{Jebari}}
	\affiliation{Univ.~Grenoble Alpes, CEA, INAC-PHELIQS, F-38000 Grenoble, France}
	\author{D. \surname{Hazra}}
	\affiliation{Univ.~Grenoble Alpes, CEA, INAC-PHELIQS, F-38000 Grenoble, France}
	\author{F. Gustavo}
	\affiliation{Univ.~Grenoble Alpes, CEA, INAC-PHELIQS, F-38000 Grenoble, France}
	\author{J.L. Thomassin}
	\affiliation{Univ.~Grenoble Alpes, CEA, INAC-PHELIQS, F-38000 Grenoble, France}
	\author{E. \surname{Dupont-Ferrier}}
	\affiliation{Institut Quantique / D\'epartement de physique, Universit\'e de Sherbrooke, Sherbrooke, QC, Canada}
	\author{F. \surname{Portier}}
	\affiliation{SPEC, CEA, CNRS, Universit\'e Paris-Saclay, CEA-Saclay 91191 Gif-sur-Yvette, France}
	\author{M. \surname{Hofheinz}}
	\email{email: max.hofheinz@usherbrooke.ca}
	\affiliation{Univ.~Grenoble Alpes, CEA, INAC-PHELIQS, F-38000 Grenoble, France}
	\affiliation{Institut Quantique / D\'epartement GEGI, Universit\'e de Sherbrooke, Sherbrooke, QC, Canada}
	
	\date{\today}
	
	\begin{abstract}
		The ability to generate single photons is not only an ubiquitous tool for scientific exploration with applications ranging from spectroscopy and metrology~\cite{Lounis2005,Eisaman2011} to quantum computing~\cite{Knill2001}, but also an important proof of the underlying quantum nature of a physical process~\cite{walls_book}. In the microwave regime, emission of anti-bunched radiation has so far relied on coherent control of Josephson qubits~\cite{Houck2007,Hoi2012,Yin2013,Pechal2014}, where precisely calibrated microwave pulses are needed, and the achievable bandwidth is limited by the anharmonicity of the qubit. Here, we demonstrate the operation of a bright on-demand source of quantum microwave radiation capable of emitting anti-bunched photons based on inelastic Cooper pair tunneling and driven by a simple DC voltage bias. It is characterized by its normalized second order correlation function of $g^{(2)}(0)\approx0.43$ corresponding to anti-bunching in the single photon regime.
		Our source can be triggered and its emission rate is tunable \textit{in situ} exceeding rates obtained with current microwave single photon sources by more than one order of magnitude.
	\end{abstract}
	\maketitle
	Josephson photonics has recently emerged as a way to directly generate and manipulate microwave frequency signals at milliKelvin temperatures without the need for complicated microwave control drives~\cite{Padurariu2012,Armour2013,Gramich2013,Dambach2015,Grimm2015,Jebari2017,Chen2014,Cassidy2017}. It relies on the phenomenon of inelastic Cooper pair tunneling where a DC voltage biased Josephson tunnel junction at zero temperature can admit a finite direct current, even if the 
	applied bias voltage is smaller than its gap voltage~\cite{Ingold1992}.
	Although the finite voltage bias makes it impossible for Cooper pairs to tunnel elastically in this regime, they can tunnel inelastically while dissipating their surplus energy into the electromagnetic environment of the junction. 
	The resulting excitations of the environment are photons at microwave frequencies~\cite{Holst1994,Hofheinz2011}. 
	This effect can be harnessed and the properties of the emitted radiation can be controlled by presenting the junction with a specifically tailored electromagnetic environment.
	
	Specifically, a Cooper pair of charge $2e$ can tunnel inelastically through a Josephson junction biased at voltage $\bias$ if the electromagnetic environment of the junction has modes that can absorb the energy difference of $2e\bias=h\Jfreq=\sum_in\sub{i}h\freq\sub{i}$, where $\Jfreq=2e\bias/h$ is the Josephson frequency and $n\sub{i}$ is the number of photons emitted into a given mode of frequency $\freq\sub{i}$. The power spectral density in units of photons of the emitted radiation at frequency $\freq$ is then given by the expression~\cite{Hofheinz2011}:
	\begin{equation}
	\gamma(\freq,\Jfreq)=\frac{\Icrit^2}{2}\frac{\operatorname{Re} Z(\freq)}{\freq}\PE(h\Jfreq-h\freq).
	\label{equ:PSD}
	\end{equation}
	Here, $\Icrit$ is the critical current of the junction, $\operatorname{Re} Z(\freq)$ is the real part of the impedance describing the electromagnetic environment and
	$\PE(h\freq)$ is the probability density for a tunneling Cooper pair
	to exchange an energy $h\freq$ with this
	environment~\cite{Ingold1992,Nazarov2013}. The latter quantity is directly proportional
	to the Cooper pair tunneling rate~$\Gamma$:
	\begin{equation}
	\Gamma(\Jfreq)=\frac{\Icrit^2}{4}\frac{h}{4e^2}\PE(h\Jfreq).
	\label{equ:Current}
	\end{equation}
	\begin{figure*}
		\includegraphics[width=0.6\textwidth]{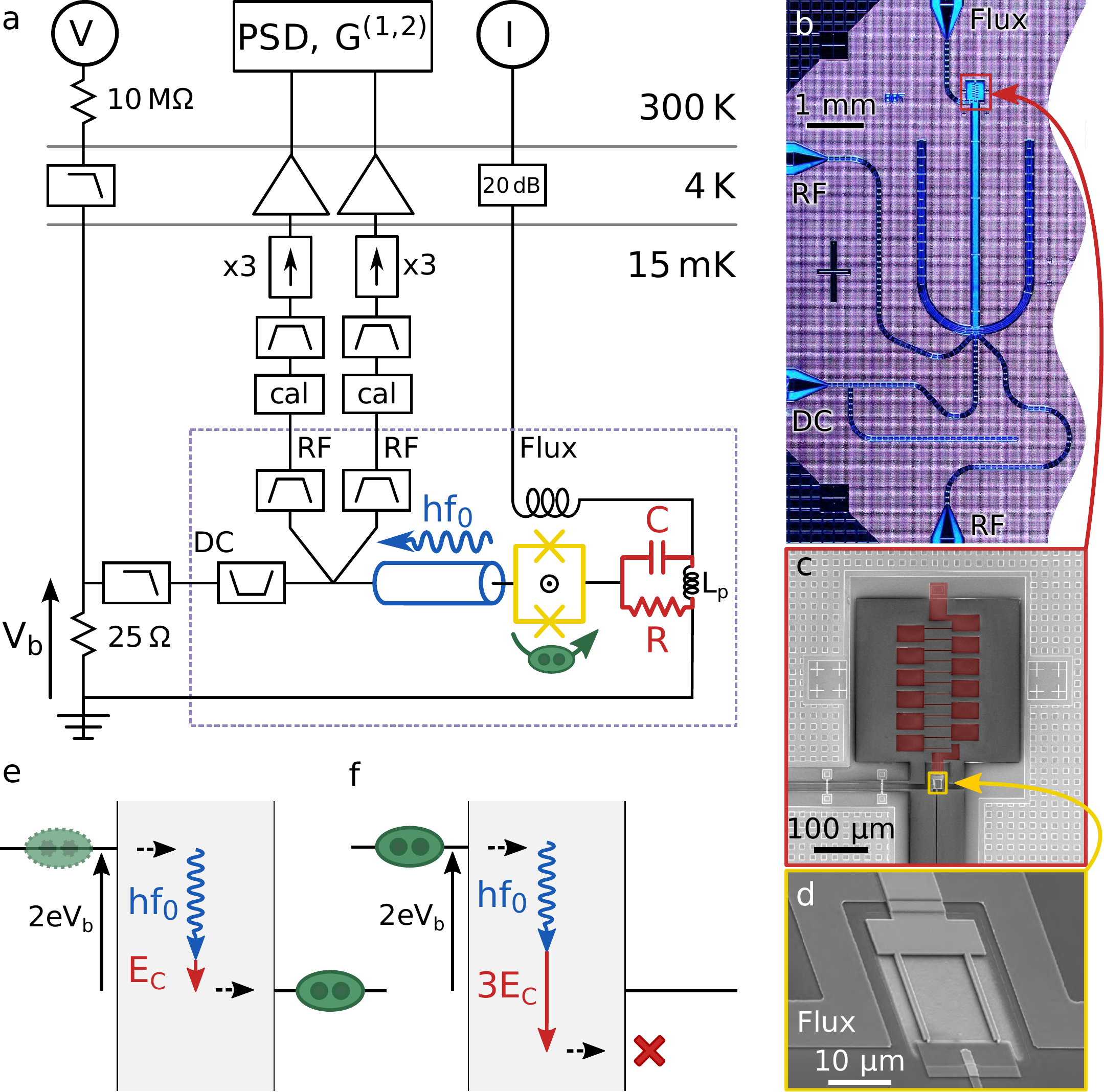}
		\caption[Sample, measurement setup and working
		principle.]{\label{fig:sample} \textbf{Sample, measurement setup,
				and working principle.}  \textbf{\circuit,} Overview of the
			sample and measurement setup. The SQUID (yellow) is connected to an RC-element (red, with parasitic inductance $L\sub{p}$) and a transmission line resonator with a fundamental mode at $\reso$ (blue). Voltage and flux biases are applied to the SQUID through a voltage divider and an on-chip fast flux line respectively. Photons escape from the resonator through an on-chip bias tee and beamsplitter into two measurement chains each containing an RF switch for calibration (marked ``cal'') as well as filters and isolators to protect the sample from amplifier noise (Supplementary Information). Power spectral density (PSD) and the
			photon correlation functions $G^{(1)}$ and $G^{(2)}$ are calculated numerically after down-mixing and digitization (Supplementary Information). 
			\textbf{\chip,} Optical image overview of
			the entire sample as indicated in the framed part of panel
			\circuit. SQUID and RC-circuit are inside the red frame. The rest of the circuit implements resonator, on-chip bias tee and beamsplitter (Supplementary Information) branching out into two high frequency (RF) outputs, and a DC input. \textbf{\rc,} Scanning electron micrograph (SEM) of the SQUID (small yellow frame) and RC-circuit (red). \textbf{\squid,} SEM of the SQUID loop consisting of two vertical NbN-MgO-NbN Josephson
			junctions. The fast flux line is visible on three sides. Where it
			is underneath the main circuit, it implements the parallel plate
			capacitor of the RC.  \textbf{\tunnel,} Schematic depiction of the
			inelastic tunneling process. Horizontal black lines represent the Cooper-pair condensates in the superconductors on either
			side of the insulating layer (gray) of the voltage biased SQUID. The
			choice of voltage bias ($2e\bias=h\reso+E\sub{C}$) is such that
			tunneling with simultaneous photon emission is possible when the
			capacitor is not already charged. \textbf{\notunnel,} Immediately
			after a tunneling event the energy necessary to charge the capacitor
			with another Cooper pair is $3E\sub{C}$, blocking further Cooper pair tunneling for a
			time $\approx RC$.}
	\end{figure*}
	
	While early experiments focused on the rate
	$\Gamma$ and the associated direct current~\cite{Delsing1989,Cleland1990,Devoret1990,Holst1994}, recent
	advances in low-noise high-frequency measurements have made it
	possible to investigate the photonic side of this energy
	transfer \cite{Hofheinz2011,Chen2014, Westig2017, Cassidy2017,
		Jebari2017}.  In all previous implementations, the tunneling events
	were happening either independently~\cite{Holst1994,Hofheinz2011} or
	through stimulated emission~\cite{Chen2014, Cassidy2017, Jebari2017}, 
	in both cases leading to classical Poisson statistics of the emitted photons.  In the present work, we demonstrate a device where the electromagnetic environment of the junction is engineered to create anti-bunching in the photon emission, thus showing quantum statistics generated through inelastic Cooper pair tunneling. 
	
	A schematic circuit-diagram of our device is shown in the bottom part of \figcite{fig:sample}{\circuit}. Its main element is a superconducting quantum
	interference device (SQUID) made of two parallel \SI{0.02}{\micro\meter\squared} NbN-MgO-NbN Josephson junctions~\cite{Grimm2017} (\figcite{fig:sample}{\squid}). We used a fast flux line to tune its critical current and thus the tunneling rate \textit{in situ}. One side of the SQUID connects to a quarter-wave transmission line resonator (blue in \figcite{fig:sample}{\circuit}), with a fundamental frequency $\reso=\SI{6}{\giga\hertz}$, which is followed by an on-chip bias tee and beamsplitter (\figcite{fig:sample}{\circuit, \chip} and Supplementary Information). The other side of the SQUID is grounded through an on-chip parallel RC-circuit (red, for details see Supplementary Information), with resistance $R$, capacitance $C$, and a small spurious inductance $L\sub{p}$.
	
	We extracted the calibrated and time-resolved auto- and cross-correlations ($G^{(1)}$, $G^{(2)}$)~\cite{daSilva2010,Eichler2012,Grimm2015} as well as the power spectral density (PSD) of the microwave radiation emitted by our device using the measurement setup depicted schematically in \figcite{fig:sample}{\circuit} and described in detail in the Supplementary Information. The PSD divided by $h\freq$ has units of photons and corresponds to the quantity computed by equation~(\ref{equ:PSD}).
	\begin{figure*}
		\includegraphics[width=\textwidth]{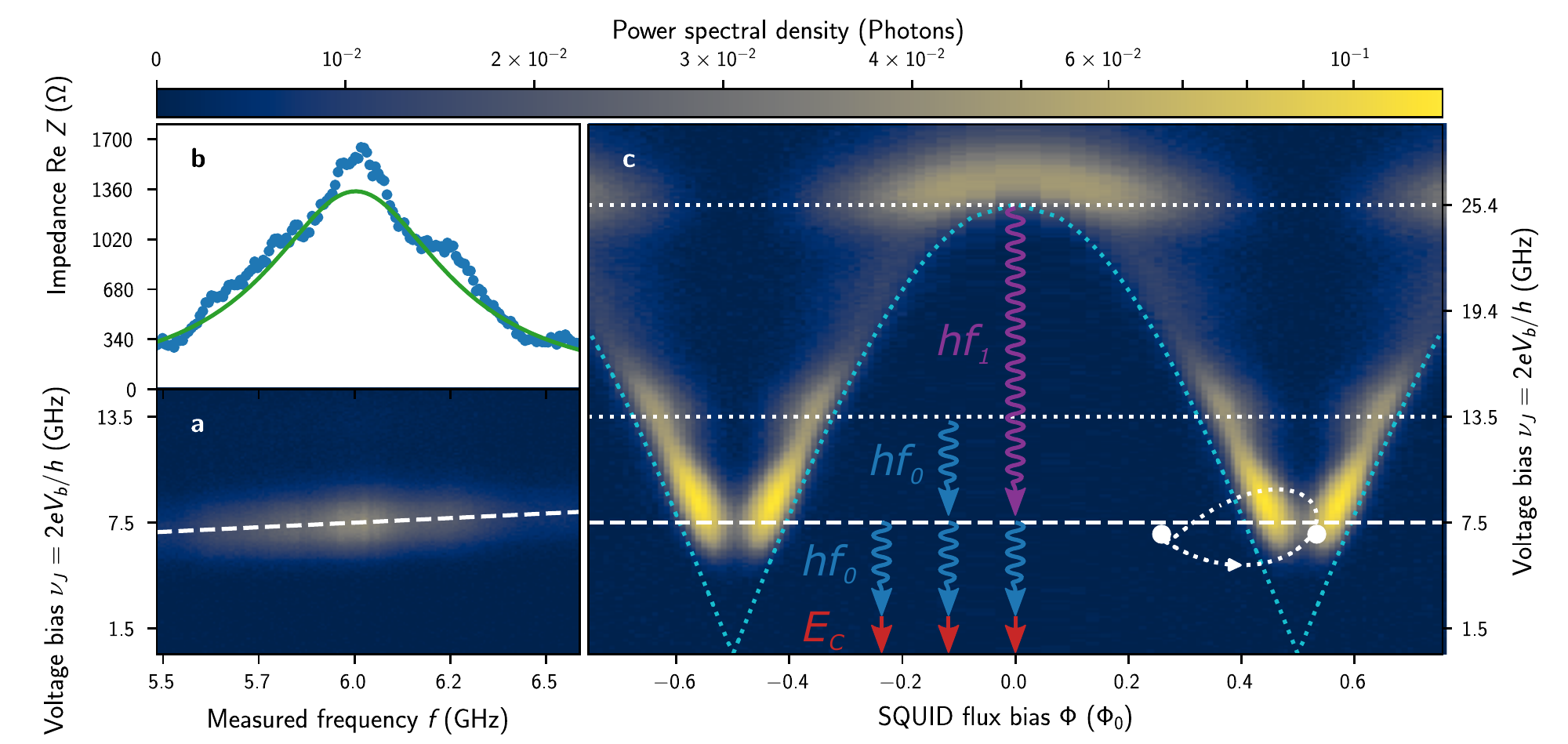}
		\caption[Power spectral density]{\label{fig:PSD}\textbf{Power
				spectral density} \textbf{\psdfreq,} Power spectral
			density in units of photons measured close to minimum $\Icrit(\Phi)$
			(SQUID flux bias of $\Phi\approx\SI{0.5}{\Phi_0}$, with
			$\Phi_0=h/2e$ the flux quantum) as a function of
			frequency and voltage bias (expressed as the Josephson frequency). The dashed line corresponds to
			the process $h\Jfreq=E\sub{C}+h\freq$, where
			$E\sub{C}\approx\EcExtracted$ is the energy required to charge the
			capacitor with a Cooper pair. The frequency axis is centered around $\reso\approx\FreqFit$,
			the resonance frequency of the transmission line resonator.
			\textbf{\impedancepeak,} Real part of the impedance as a
			function of the voltage bias directly extracted from the
			data shown in panel \psdfreq~(blue dots) and plotted from a fit of our circuit model to the data (solid
			green line).  \textbf{\psdflux,} Power spectral density in
			units of photons at \SI{6}{\giga \hertz} as a function of voltage bias and flux
			bias in units of flux quanta in the SQUID loop. The horizontal lines are drawn using the values of $E\sub{C}$ and $\reso$ extracted from panel a and
			match the observed features. The underlying emission process is sketched for each line with the dashed one corresponding to the
			first order process $h\Jfreq=E\sub{C}+h\reso$ and the dotted ones
			corresponding to second order processes
			$h\Jfreq=E\sub{C}+2h\reso$ and $h\Jfreq=E\sub{C}+h\reso+h\freq\sub{1}$, where $\freq\sub{1}$ is the frequency of the next higher resonator mode. The turquoise dotted line delimits the region in which no voltage bias can
			be applied because the SQUID is latching to its current branch (see text). Two filled white circles mark the starting and end-point of the
			flux pulse giving the results of \figcite{fig:pulsed}{} and
			our estimate of the actual trajectory is sketched in white.}
	\end{figure*}
	
	The underlying principle of our source is that one tunneling event necessarily acts back onto the next one in
	order to create anti-correlations in the Cooper-pair current, leading to anti-bunching in the photon emission. On timescales shorter than $RC$, a tunneling Cooper-pair has to charge the island formed between capacitor and SQUID. To emit a photon, the voltage bias must be chosen such that $h\Jfreq=h\reso+E\sub{C}$, where $E\sub{C}=(2e)^{2}/2C$ is
	the charging energy of the island (\figcite{fig:sample}{\tunnel}). However, immediately after a
	first tunneling event the energy necessary to add another Cooper-pair to the island is $(2\cdot 2e)^{2}/2C-(2e)^{2}/2C=3E\sub{C}$.
	This makes the voltage bias insufficient for further tunneling (\figcite{fig:sample}{\notunnel}). A second Cooper pair can tunnel and emit a photon only after a time $RC$, when the island is discharged, leading to the desired anti-bunching. This picture is valid if the charging energy associated with one Cooper pair is large compared to the thermal energy fluctuations, i.e. $E\sub{C}\gg k\sub{B}T\sub{eff}$, as well as the lifetime broadening of the charging energy, i.e. $E\sub{C}\gg \hbar/(2 RC)$, or equivalently $R\gg h/(4e^{2})$. Here, $T\sub{eff}$ is the effective temperature of the low-frequency electromagnetic environment.
	
	We now verify these conditions and extract the device parameters. To this end, we analyze the PSD in \figcite{fig:PSD}{\psdfreq} taken at a flux bias $\Phi \approx 0.5\Phi_0$, where $\Phi_0=\frac{h}{2e}$ is the superconducting flux quantum. There, $\Icrit(\Phi)$ and thus emission rates are low so that tunneling events can be
	considered independent and equations~(\ref{equ:PSD}) and
	(\ref{equ:Current}) are valid~\cite{Ingold1992}. For this flux bias, photon emission is maximal at $\Jfreq = \SI{7.5}{\giga\hertz}$ and $\freq = \reso=\FreqFit$. The
	difference between the energy $h\Jfreq$ provided by a tunneling Cooper
	pair and the energy $h\freq$ of the detected photons is
	the charging energy $E\sub{C}\approx h\times\EcExtracted\approx k\sub{B} \times \SI{72}{\milli \kelvin}$ ($C\approx\SI{51}{\femto \farad}$). Further analysis of this PSD (Supplementary information) yields the real part of the resonator impedance (blue dots in \figcite{fig:PSD}{\impedancepeak})
	and the effective temperature of the low-frequency electromagnetic environment $T\sub{eff} \approx\SI{21}{\milli \kelvin}$. The resistance $R\approx\RExtracted$ was determined from an independent DC measurement (Supplementary Information). Additionally, we verify the accuracy of the electrical model presented in \figcite{fig:sample}{\circuit} by using it to fit equation~(\ref{equ:PSD}) to the PSD. The real part of the resonator impedance given by the resulting circuit parameters is shown as a solid green line in \figcite{fig:PSD}{\impedancepeak} (for detailed parameters see Supplementary Information). It has a full width at half maximum of \SI{575}{\mega \hertz} and agrees well with the measured data. The $\approx\SI{200}{\mega\hertz}$-periodic impedance modulations are due to reflections in our output lines, which are not included in the model.
	
	In \figcite{fig:PSD}{\psdflux} we now explore the behavior of the device
	when the critical current is increased to values relevant for
	operation, where Cooper pair tunneling events cannot be considered
	independent any more. To do so, we measure the power spectral density
	at its maximal value in frequency (at $\reso$) as a function of $\Jfreq$ and the flux bias $\Phi$. The brightest
	features appear around $\Jfreq = \SI{7.5}{\giga\hertz}$,
	corresponding to the desired process. At slightly lower flux bias
	additional features appear at $\Jfreq = \SI{13.5}{\giga\hertz}$. They
	originate from the emission of two photons per
	tunneling Cooper pair into the mode at $\reso$. When the critical current is maximal ($\Phi = 0$), another feature becomes visible at $\Jfreq =
	\SI{25.4}{\giga\hertz}$. It corresponds to processes where one Cooper
	pair emits one photon into mode $\reso$ and one into the next higher mode of the quarter-wave transmission line resonator at $\freq_1\approx3\reso$.
	
	Strikingly, the processes at $\SI{7.5}{\giga\hertz}$ and
	$\SI{13.5}{\giga\hertz}$ disappear around $\Phi=0$ where one would
	expect the rates to be highest. We instead observe a dark zone delimited by the dotted parabola. In this region the critical
	current is high enough for the SQUID to get trapped
	in a Bloch oscillation regime~\cite{Negri2012,Likharev1988,vora2017,corlevi2006}, a rudiment of the zero-voltage state observed in
	larger Josephson junctions. In this state the voltage drops mostly
	over the RC-element, decreasing the voltage difference across the SQUID below the threshold for photon emission at $\reso$. This interpretation is confirmed by an independent measurement of
	the resonator frequency showing flux-tunability in the region delimited by the dotted
	line (Supplementary Information).
	
	We now focus on the key question
	of this work and investigate the statistics of the radiation emitted
	by the device. To do so, we measure its normalized second-order
	correlation function $g^{(2)}(t,\tau)$ given by~\cite{Glauber1963}
	\begin{equation}
	g^{(2)}(t,\tau)=
	\frac{G^{(2)}(t,\tau)}{G^{(1)}(t,0)G^{(1)}(t+\tau,0)}.
	\label{eq:2}
	\end{equation}
	In the above expression, $G^{(2)}(t,\tau)$ is the unnormalized second-order correlation 
	function dependent on a time $t$ with respect to a reference and on the time delay $\tau$, defined as:
	\begin{equation}
	G^{(2)}(t,\tau)=\left<
	\hat{a}^{\dagger}_{\rm out}(t)
	\hat{a}^{\dagger}_{\rm out}(t+\tau)
	\hat{a}^{\phantom{\dagger}}_{\rm out}(t+\tau)
	\hat{a}^{\phantom{\dagger}}_{\rm out}(t)\right>.
	\label{eq:3}
	\end{equation}
	The operators $\hat{a}^{\phantom{\dagger}}_{\rm out}$ and
	$\hat{a}^{\dagger}_{\rm out}$ are the annihilation and creation operators
	of the outgoing field in the transmission line. The denominator of the right-hand side of \eqcite{eq:2} is a normalization factor dependent on the first-order correlation function
	\begin{equation}	
	G^{(1)}(t,\tau)=\left<\hat{a}^{\dagger}_{\rm out}(t)\hat{a}^{\phantom{\dagger}}_{\rm out}
	(t+\tau)\right>.
	\label{eq:4}
	\end{equation}
	$G^{(1)}(t,0)$ gives the photon emission rate of our device. 
	In the absence of a well defined time reference, an average over $t$ is performed on 
	$G^{(1)}(t,\tau)$ and $G^{(2)}(t,\tau)$ yielding 
	$G^{(2)}(\tau)$, $G^{(1)}(\tau)$ and $g^{(2)}(\tau)$ \cite{Glauber1963}.
	
	\figcite{fig:free}{a} shows the total photon emission rate $G^{(1)}(0)$ of our sample in a region around the one-photon peak visible in Fig.~\ref{fig:PSD}{\psdflux}. It is measured by integrating the PSD over frequency between \SI{4.22}{\giga \hertz} and \SI{8}{\giga \hertz} (Supplementary Information). We have evaluated $g^{\left(2\right)}\left(\tau\right)$ at different points along two lines of constant voltage (flux) bias as indicated by plus (cross) symbols, corresponding to the curves shown in \figcite{fig:free}{b} (\figcite{fig:free}{c}).
	
	\begin{figure}[h]
		\includegraphics[width=\columnwidth]{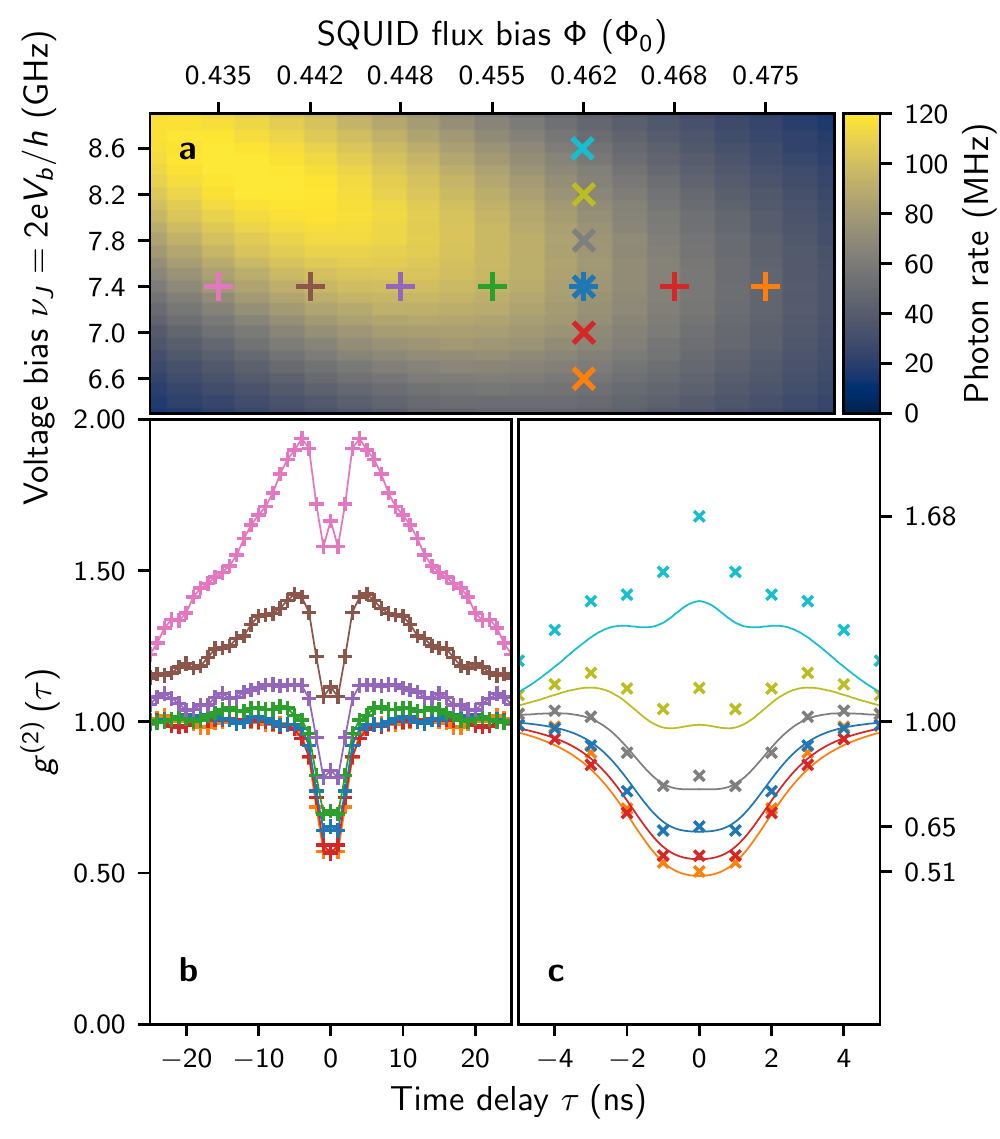}
		\caption[Free-running operation]{\label{fig:free}\textbf{Emission rate and statistics}
			\textbf{a, }Photon emission rate measured around the one-photon peak from \SI{4.22}{\giga \hertz} to \SI{8}{\giga \hertz}. 
			Symbols mark the points where the data shown in panels b and c was taken. 		
			\textbf{b, }Second order correlation functions at voltage bias $\Jfreq = \SI{7.4}{\giga\hertz}$ for different flux biases. Lines are a guide for the eye. For clarity, error bars are omitted here and in panel c ($3\sigma\approx0.1$, see Supplementary Information). Curves from top to bottom correspond to the {+} symbols in panel a from left to right.	
			At low flux values the SQUID switches randomly between its current and voltage branches, which is visible as a 
			reduced photon emission rate in panel a. This leads to an overall bunching peak superimposed with the anti-bunching dip. 
			\textbf{c, }Second order correlation functions at flux bias $\Phi = \SI{0.462}{\Phi\sub{0}}$ for different voltage biases. Curves from top to bottom correspond to the $\times$ symbols in panel a from top to bottom.
			Higher voltage biases enable multiple tunneling events and thus bunching.
			Solid lines are numerical simulations following reference~\onlinecite{Leppakangas2015} (see text).}
	\end{figure}
	
	Close to $\Phi\approx\SI{0.5}{\Phi\sub{0}}$, the $g^{(2)}(\tau)$ function shows a marked dip down to approximately 0.5 at $\tau = 0$ and is close to the expected value of 1 elsewhere, indicating that our sample indeed produces strongly anti-bunched photons. \figcite{fig:free}{b} shows that, as we approach lower flux biases to increase the critical current, the sharp dip close to $\tau=0$ remains, but a broad bunching peak develops around it. We attribute this broad peak to random jumps between the bright voltage state of the junction and the dark zero-voltage state discussed above. This is consistent with the reduction in emission rate towards the lower left corner of \figcite{fig:free}{a}. The persistence of the sharp dip at $\tau=0$ demonstrates that the intended blocking mechanism remains functional, despite this effect. 
	
	\figcite{fig:free}{c} shows the sensitivity of the anti-bunching to precise adjustment of the bias voltage. Here we chose the flux bias where we achieve the highest photon rate while bunching due to jumps to the dark state remains negligible: $\Phi\approx\SI{0.462}{\Phi\sub{0}}$. When biased above \SI{8}{\giga\hertz}, our sample emits bunched light ($g^{(2)}(0)>1$). This bunching effect can be understood
	by considering the energy diagrams in \figcite{fig:sample}{\tunnel, \notunnel} at higher voltage biases: Even though the voltage is initially too high for the resonance condition to be fulfilled, thermal fluctuations still occasionally allow Cooper pairs to tunnel. When a first Cooper pair does tunnel, the RC-circuit is charged, and a second one can more easily follow, leading to bunching. As the bias decreases, so does the value of the second order
	correlation function, down to $g^{(2)}(0)\approx0.5\pm0.1(\pm3\sigma)$. The fact that the lowest value of
	$g^{(2)}(0)$ is reached below the maximum of the peak is expected:
	residual double emission events caused by thermal fluctuations
	of the bias voltage are further suppressed. The blue curve is taken at the maximum emission rate of \SI{77}{\mega\hertz} ($\Jfreq=\SI{7.4}{\giga \hertz}$) along this cut. 
	At this point we measure $g^{(2)}(0)\approx 0.65\pm0.06$. We conclude that anti-bunching is robust as long as the bias voltage is kept at the nominal resonant value or below, but is rapidly transformed into bunching above. 
	
	The solid lines in \figcite{fig:free}{c} are numerical calculations up to fourth order in the critical current \cite{Leppakangas2016, Leppakangas2015} using the electrical model of \figcite{fig:sample}{\circuit} with the extracted device parameters and an effective temperature of $T\sub{eff} = \SI{40}{mK}$. They reproduce well the observed anti-bunching signatures at the lowest bias voltages, including bumps in $g^{(2)}$ at approximately \SI{2}{\nano\second} but fail to fully explain the bunching signatures for the highest bias voltages. In addition, $T\sub{eff}$ used here to reproduce the data is significantly higher than the temperature extracted from the data in \figcite{fig:PSD}{\psdfreq} at low critical current and at low bias voltage. This discrepancy could indicate that the resistor of the RC element heats up more than expected at the higher photon fluxes and higher voltages used here. Another explanation could be that correlations between more than two Cooper pairs are relevant and that calculations have to be performed beyond fourth order in the critical current, which significantly increases the computational effort and is left for future work.
	
	So far we have focused on the free-running mode of operation, where latching to the dark state prevents us from reaching
	higher emission rates. However, as we now
	show, we can make use of this effect to produce photons on demand through the
	flux-pulsing scheme indicated in white on \figcite{fig:PSD}{\psdflux}. For this, the voltage bias is set to its nominal value ($\Jfreq=
	\SI{7.4}{GHz}$). We start out at a flux bias well in the
	dark region (left white dot), where no photon emission occurs in the
	stationary regime. We then apply a flux pulse which frustrates the SQUID (right white dot), suppressing its critical current and resetting it to the voltage branch. At this bias point photon emission is unlikely. Returning to the initial point in the dark region, removes the frustration and allows a photon to be emitted (and the capacitor to be charged). Then, the system is again trapped in the dark state limiting photon emission to at most one per cycle.
	
	\figcite{fig:pulsed}{} shows the data obtained when Gaussian flux bias pulses with FWHM of $\SI{1.5}{\nano\second}$ are applied every \SI{6}{\nano\second}. Different pulse durations (within a factor 2) and lower repetition rates do not affect the results significantly. Higher repetition rates, however, cause photon pulses to overlap.
	
	In \figcite{fig:pulsed}{a} we first explore
	the photon emission rate $G^{(1)}(t,0)$ as a function of time $t$
	with respect to the pulse. One averaged measurement yields the values marked by blue
	dots, which are separated by the sampling period of
	\SI{1}{\nano\second} of our measurement. By shifting
	the time delay between pulse-generation and the beginning of the
	sampling window, the photon emission rate can be resolved below the
	sampling period yielding the orange curve in \figcite{fig:pulsed}{a}.  The
	zero on the time axis matches
	the time $t=0$ at which the measurements in \figcite{fig:pulsed}{b, c} were
	taken. The width of this peak is due to a combination of the uncertainty in
	the photon generation time (jitter) and the resonator decay time. 
	\begin{figure}[h]
		\includegraphics[width=\columnwidth]{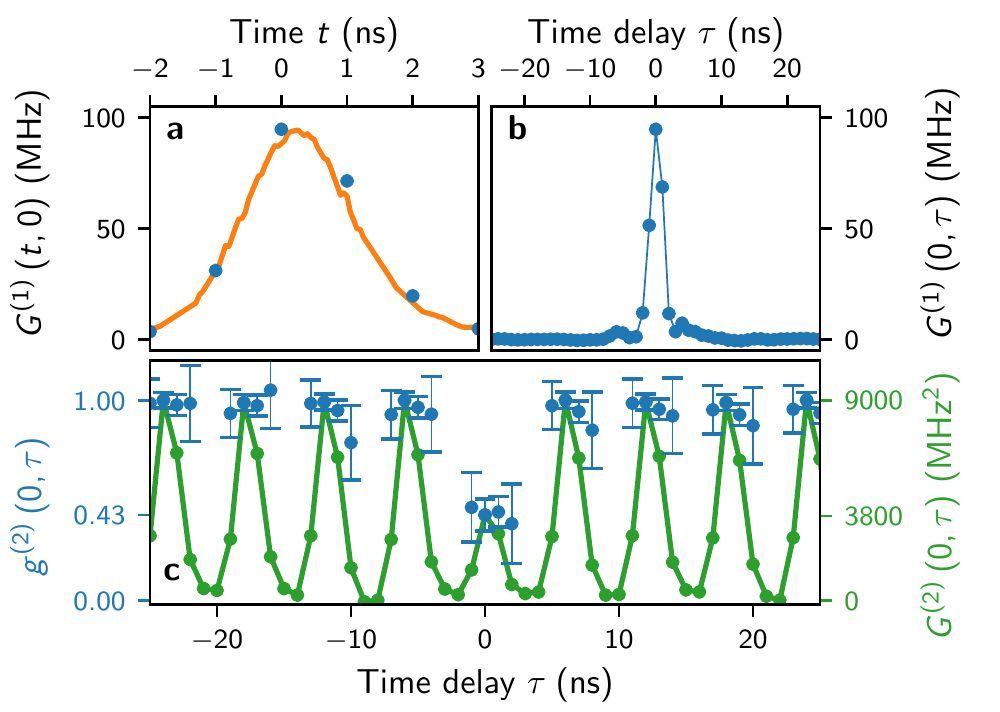}
		\caption[Flux-pulsed on-demand
		operation]{\label{fig:pulsed}\textbf{Flux-pulsed on-demand
				operation} \textbf{a, }Photon emission rate as a function
			of time.  Blue dots are the rates acquired during one
			averaged measurement, the orange line is obtained by
			repeating the measurement with different time delays of the
			flux-pulse. The
			shape of the peak is slightly skewed due to the
			finite time the photon spends in the resonator. The statistical errors in this panel and the next are small compared to the data points (Supplementary Information) \textbf{b,
			}First order correlation function at $t=0$, as a function of
			the time delay $\tau$ (the solid line is a guide for the eye). As the actual time of emission is
			slightly after $t=0$, the peak is centered at slightly positive $\tau$.  \textbf{c, }Second
			order correlation function at $t=0$, as a function of the
			time delay $\tau$.  The green dots correspond to the
			unnormalized second order correlation function (right
			ordinate) and display periodic peaks every $\SI{6}{\nano\second}$ given by the period
			of the applied flux pulse (the line is a guide for the eye). Blue dots with error bars ($\pm3\sigma$) show the normalized
			correlation function (left ordinate).}
	\end{figure}
	
	The complementary measurement of $G^{(1)}(0,\tau)$ (\figcite{fig:pulsed}{b}) characterizes the first order coherence
	of the source.  The width of the peak is very close to the peak in
	$G^{(1)}(t,0)$ in \figcite{fig:pulsed}{a}, indicating that the
	broadening of $G^{(1)}(t,0)$ is not dominated by jitter and that
	photons generated by our source may be
	indistinguishable. Also note that
	$G^{(1)}(0,\SI{6}{\nano\second})\approx 0$, indicating that there is
	no coherence between successive pulses as expected for single
	photons.
	
	In the unnormalized second order correlation function $G^{(2)}(0,\tau)$ this picture is reversed (\figcite{fig:pulsed}{c}). Peaks appear every \SI{6}{\nano\second}, equal to the period of the applied flux-pulse, proving the on-demand aspect of our source. Note that the peak at $\tau=0$ is significantly lower than the others, indicating the anti-bunched character of the emitted radiation.
	
	In order to quantify the anti-bunching we compute the normalized second order correlation function (blue dots) according to \eqcite{eq:2}. This is only done for times with high enough 
	emission rate $G^{(1)}(0+\tau,0)$, in order to avoid divergence of the associated errors. 
	For time-differences $\tau=n\times\SI{6}{\nano \second}$ with $n\neq0$, we obtain $g^{(2)} \approx 1$, indicating that photons 
	from two different pulses are independent. 
	At zero time-delay, however, $g^{(2)}(0,0)=0.43\pm0.08(\pm3\sigma)$: the photons are strongly anti-bunched, 
	in agreement with the mechanism presented in \figcite{fig:sample}{\tunnel, \notunnel.}
	
	We achieve here stronger anti-bunching than in the free-running case, likely due to the additional blocking effect given by the latching to the zero-voltage state in the dark region. At the same time, the pulsing scheme allows us to maintain very good quantum efficiency and photon flux ($0.2$ photons per pulse), because of the high emission rates at low flux bias. This makes it likely for a tunnel event to happen during each cycle even for very short flux pulses. We attribute the residual $g^{(2)}(0,0)$ mainly to the low charging energy of our RC-circuit and its relatively low time constant ($RC=\SI{1.64}{\nano \second}$), only slightly larger than the decay time of the resonator $\tau=\SI{0.28}{\nano \second}$. The latter is limited by practical considerations such as the instantaneous bandwidth of our measurement setup (Supplementary Information). In addition, a parasitic coupling between the flux bias and the RC-circuit, described by the parasitic inductance $L\sub{p}$ in Fig.~\ref{fig:sample}\circuit, causes the flux pulse to modulate the voltage across the junction. The dashed white line in \figcite{fig:PSD}{\psdflux} indicates the resulting expected trajectory in parameter space, which may not be ideal. Further optimization of this trajectory and modification of the coupling or application of simultaneous voltage and flux pulses could lead to better anti-bunching. Moreover, this type of device could possibly be optimized as a source of on-demand multi-photon Fock states by addressing the processes appearing at higher bias voltages.

	In conclusion, we have demonstrated a Josephson photonics device producing strongly anti-bunched microwave
	radiation. In doing so we have shown that quantum statistics can be generated from inelastic Cooper pair tunneling. By
	modulating the effective critical current of the SQUID, using fast flux
	pulses and locking to a dark state after photon emission, we
	can produce anti-bunched photons on-demand at very high rates. Increasing the
	charging energy and the RC time, or replacing the RC-circuit by a high
	impedance resonant mode, should allow for significant improvements
	of anti-bunching and quantum efficiency in future iterations of the device. We expect that it can be optimized to be an on-demand single photon source with near
	unity quantum efficiency. Such a source could then be used for quantum
	metrology, quantum computation with propagating photons, or quantum
	measurements in cases where the shot noise of coherent light needs to
	be avoided.
	\newline
	\begin{acknowledgments}
		We acknowledge fruitful discussions with the Quantronics group, B. Kubala and J. Ankerhold, Fran\c{c}ois Lefloch and Thierry Chevolleau, as well as financial support from the Grenoble Nanosciences Foundation (grant WiQOJo), the European Research Council (starting grant 278203 WiQOJo) and the Agence Nationale de la Recherche (grant JosePhSCharLi). 
	\end{acknowledgments}
	
	\bibliography{SPS}
	
	\begin{ac}
		AG and MH designed the sample and built the initial experiment based on an initial idea by MH and discussions with FP and EDF. AG, SJ, DH, FG and JLT brought up the fabrication process. AG, FG and JLT built the sample. AG performed initial measurements and data-analysis. FB and RA improved the measurement setup, datataking and calibration routines, took final data and performed final data analysis. JL provided theory support. AG wrote the paper with figures from FB and input from all authors.
	\end{ac}

\newpage
\onecolumngrid

\begin{center}
	\textsc{\Large{Supplementary Information}}
\end{center}

\section{Sample fabrication}
The samples were fabricated in a multi-layered process on a Si($\SI{500}{\micro \meter}$):$\textrm{SiO}_{2}$($\SI{500}{\nano\meter}$) substrate. We first sputtered an MgO($\SI{20}{\nano \meter}$) buffer layer and an NbN($\SI{80}{\nano\meter}$)-MgO($\SI{4}{\nano\meter}$)-NbN($\SI{200}{\nano \meter}$) tri-layer, into which we subsequently etched steps defined by a combined optical (OL) and electron-beam lithography (EBL) process. In this way we could simultaneously define small features (Josephson junctions) and large features (coplanar waveguides) for many samples on a 4 inch wafer. After this, a $\SI{500}{\nano\meter}$ thick dielectric layer ($\textrm{Si}_{3}\textrm{N}_{4}$) was deposited conformally by chemical vapor deposition (CVD) and then, following an OL step, etched directionally to leave spacers isolating the tri-layer sidewalls. A counter electrode NbN layer ($\SI{300}{\nano\meter}$) was then deposited and, after another combined OL and EBL step, etched where needed to define junctions. More details on this process and an in-depth characterization of the fabricated junctions and the different superconducting thin films can be found in reference~\onlinecite{Grimm2017}. On the device at hand the SQUID is formed by two \SI{0.02}{\micro\meter\squared} Josephson junctions and has a total room temperature resistance of $\approx\SI{230}{\kilo \ohm}$. The gap voltage extracted from a measurement of the current-voltage characteristic (not shown) is $V\sub{gap}=\SI{4.8}{\milli \volt}$.

The $\textrm{Si}_{3}\textrm{N}_{4}$ layer also serves as the dielectric in a parallel plate capacitor between the tri-layer and the counter-electrode, which implements the capacitive part of the $RC$ element. It is visible in Fig.~\overview\rc~and \squid. The top NbN electrode passes over the grounded flux-bias line (implemented using the tri-layer) and contacts the resistor on the other side.

\begin{figure}[h!]
	\centering
	\includegraphics[width=0.5\linewidth]{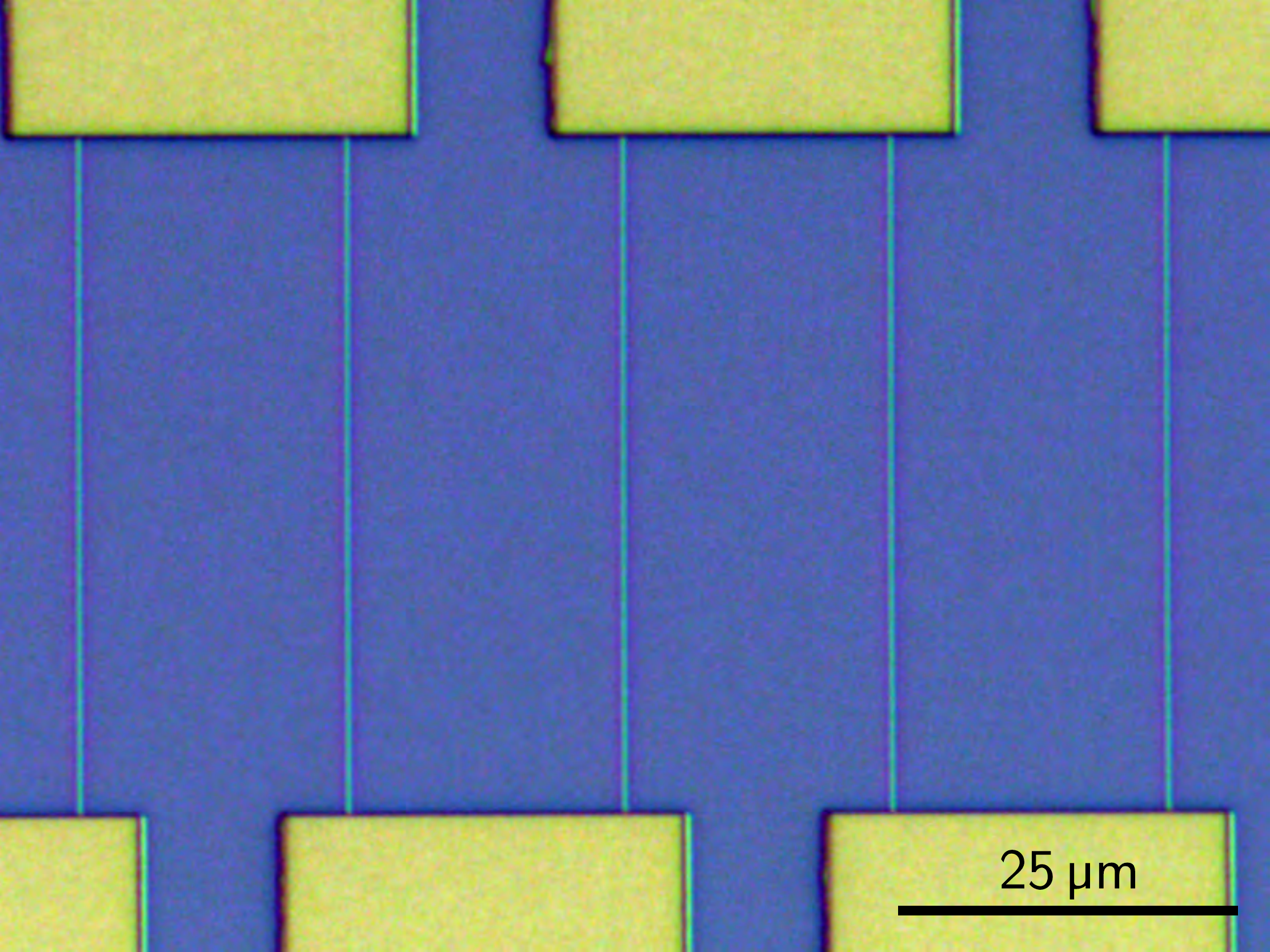}
	
	\caption{Optical micrograph of part of an on-chip Cr resistor on a device similar to the one described in the main text showing thin resistive lines and large cooling pads.}
	\label{fig:Cr}
\end{figure}

We fabricated the resistive part of the $RC$ using chromium, which was chosen for its relatively high resistivity allowing us to limit stray capacitances by keeping the resistor short. We implemented it in a final fabrication step directly on the MgO buffer layer in an area where all other layers had been etched away. For reasons detailed in Sec.~\ref{sec:heating} this element alternates thin resistive lines ($\SI{15}{\nano \meter}$) with much thicker cooling pads (\SI{100}{\nano \meter}). Both were deposited in the same fabrication step by using angle evaporation and liftoff. For this, an EBL step (ZENON-ZEP520A and AR-300-70) first defined $\SI{300}{\nano \meter}$ wide trenches connecting much larger rectangles (see Fig.~\overview\rc~and Fig.~\ref{fig:Cr}). The thin lines were deposited by evaporating perpendicularly to the wafer plane into the narrow trenches. Afterwards, the direction of evaporation was tilted so that its angle with respect to the plane of the wafer was $\SI{35}{\degree}$ and the angle between its projection onto the plane of the wafer and the trenches was $\SI{90}{\degree}$. In this orientation  $\SI{174}{\nano \meter}$ of chromium (vertical thickness: $\approx\SI{100}{\nano \meter}$) were deposited. Since the resist layer was $>\SI{300}{\nano \meter}$ thick, the Cr from this evaporation step did not reach the bottom of the trenches and was removed during lift-off.

\section{Sample design}
\label{design}
\subsection{Coplanar waveguides}
\begin{figure}
	\centering
	\includegraphics[width=0.5\linewidth]{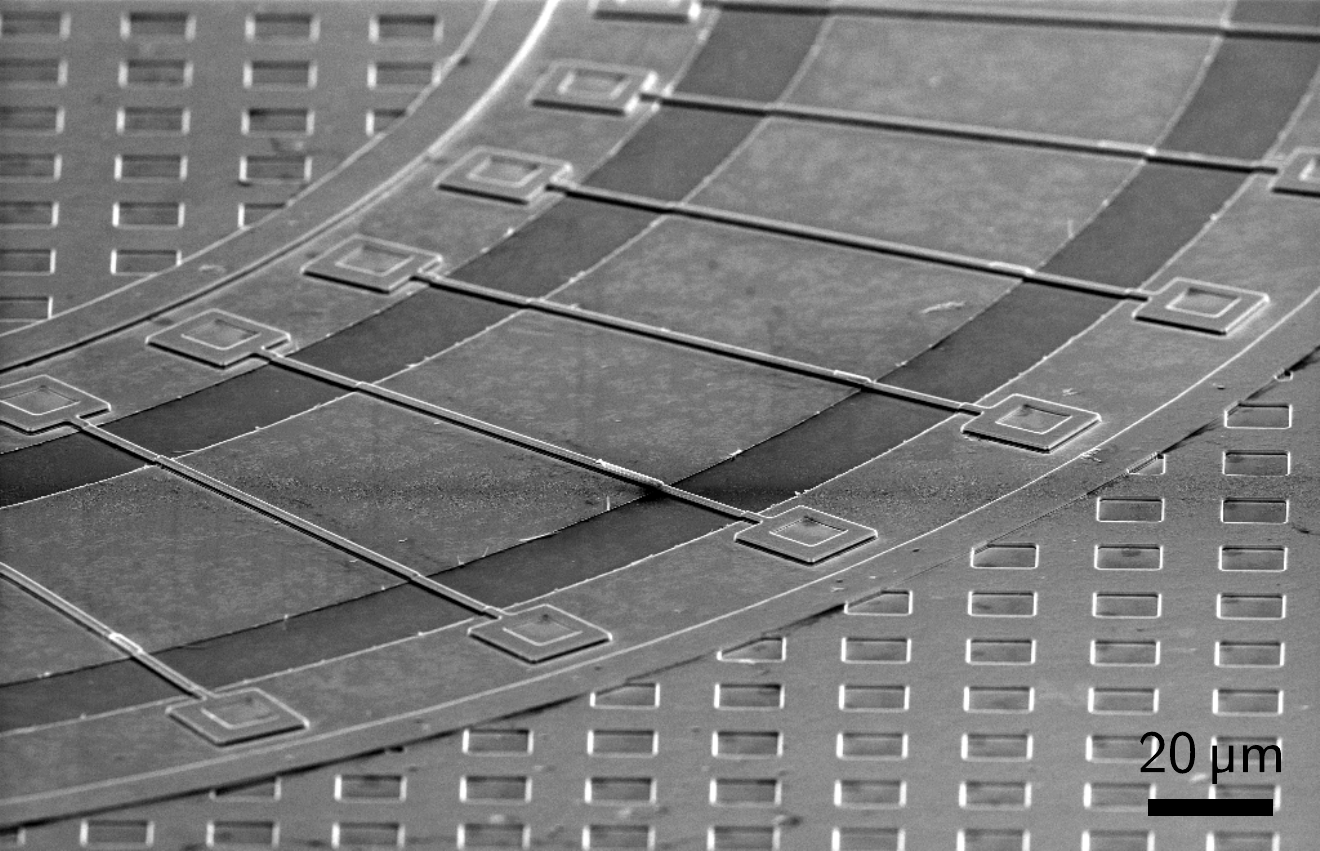}
	\caption{SEM micrograph of a coplanar waveguide with ground bridges connecting the ground planes on either side of the inner conductor. The squares on the surrounding ground planes are regions where the counter-electrode NbN layer has been etched away to create flux traps and prevent macro-loading effects during dry-etching.}
	\label{fig:CPW}
\end{figure}

The basic coplanar waveguide (CPW) structures (center conductor and ground planes) were defined in the first optical lithography step and then etched into the tri-layer. We implemented ground-bridges over the CPWs by opening up vias through the $\textrm{Si}_{3}\textrm{N}_{4}$ above the ground planes on either side of the center conductor and connecting them with the counter-electrode NbN-layer deposited in a later fabrication step (Fig.~\ref{fig:CPW}).

In order to determine the required geometries for CPWs of different characteristic impedances we used a Python-based 2D boundary element simulation program to compute capacitance and inductance per unit length. We took into account the kinetic inductance of the different NbN thin films, which we estimated from independent measurements~\cite{Grimm2017, Grimm2015}. It was necessary to correct for the presence of the ground-bridges, since the capacitance to ground of these sections is greatly enhanced. We performed a separate simulation for this specific geometry and then averaged the two types over the length of the line using effective widths for the bridges that take into account their lateral capacitance. This was done by approximating them to microstrip lines across the central conductor and comparing the resulting capacitance to that of a parallel plate capacitor. Table~\ref{tab:CPWparams} shows the simulated parameters for different types of CPW elements on the device at hand.

\begin{table}[h!]
	\centering
	\begin{tabular}{c|c|c|c|c|c|c}
		$Z_{0}$ ($\si{\ohm}$) 
		& $C$ ($\si{\pico \farad \per \meter}$)
		& $L$ ($\si{\nano \henry \per \meter}$)& $L\sub{kin}$ ($\si{\percent}$) 
		& $S$ ($\si{\micro \meter}$) & $W$ ($\si{\micro \meter}$)
		& $d\sub{bridge}$ ($\si{\micro \meter}$)\\ \hline
		146  & 64 & 1361 & 26 & 2 & 49 & 1000 \\ \hline
		90  & 114 & 898 & 33 & 3 & 8.5 & 100 \\ \hline
		70  & 140 & 719 & 29 & 5 & 7.5 & 100 \\ \hline
		50  & 204 & 512 & 27 & 10 & 5 & 100 \\
	\end{tabular}  
	\caption{Some relevant parameters of the CPW geometries used in this work including the characteristic impedance ($Z_{0}$), the capacitance ($C$) and inductance ($L$) per unit length and the percentage of the total inductance due to kinetic inductance ($L\sub{kin}$). $S$ is the width of the center conductor and $W$ is the width of the gap separating the center conductor from the ground plane on either side. The ground bridges have a cross-sectional width of $\SI{2}{\micro \meter}$ and the distance between the centers of two consecutive bridges along the CPW is given by $d\sub{bridge}$.}
	\label{tab:CPWparams}
\end{table}

\subsection{Resonator}
The resonator is formed by a $\lambda/4$ segment of CPW with a 
characteristic impedance of $Z_{\rm R} = \SI{146}{\ohm}$ (design value) 
connected on one side to the junction (point 0 in 
Fig.~\ref{fig:beamsplitter}a). When the other end is short-circuited, 
this CPW segment forms a resonator with resonances at $f_n = (2n+1) 
\frac{c}{4 l}$ where $c$ is the phase velocity in the CPW and $l$ its 
length. These resonances have a characteristic impedance of $Z_n = 
\frac{4}{\pi (2n+1)} Z_{\rm R}$. In the present case the junction capacitance slightly lowers $f_n$ and $Z_n$ and the fact that the other end of 
the CPW is not connected to a short circuit but to a beamsplitter (see 
below) at point C in Fig.~\ref{fig:beamsplitter}a gives rise to a finite quality factor. Close to resonance 
this beamsplitter loads the resonator CPW with an effective impedance of 
$Z_{\rm C} = \SI{12}{\ohm}$ leading to a quality factor of the resonator of $Q_n = 
\frac{\pi}{4}\frac{Z_R}{Z_{\rm C}} \approx \frac{9.6}{2n + 1}$, 
corresponding to a width of approximately \SI{600}{MHz}.

\subsection{On-chip beamsplitter and bias tee}

\begin{figure}
	\centering
	\includegraphics[width=1.0\linewidth]{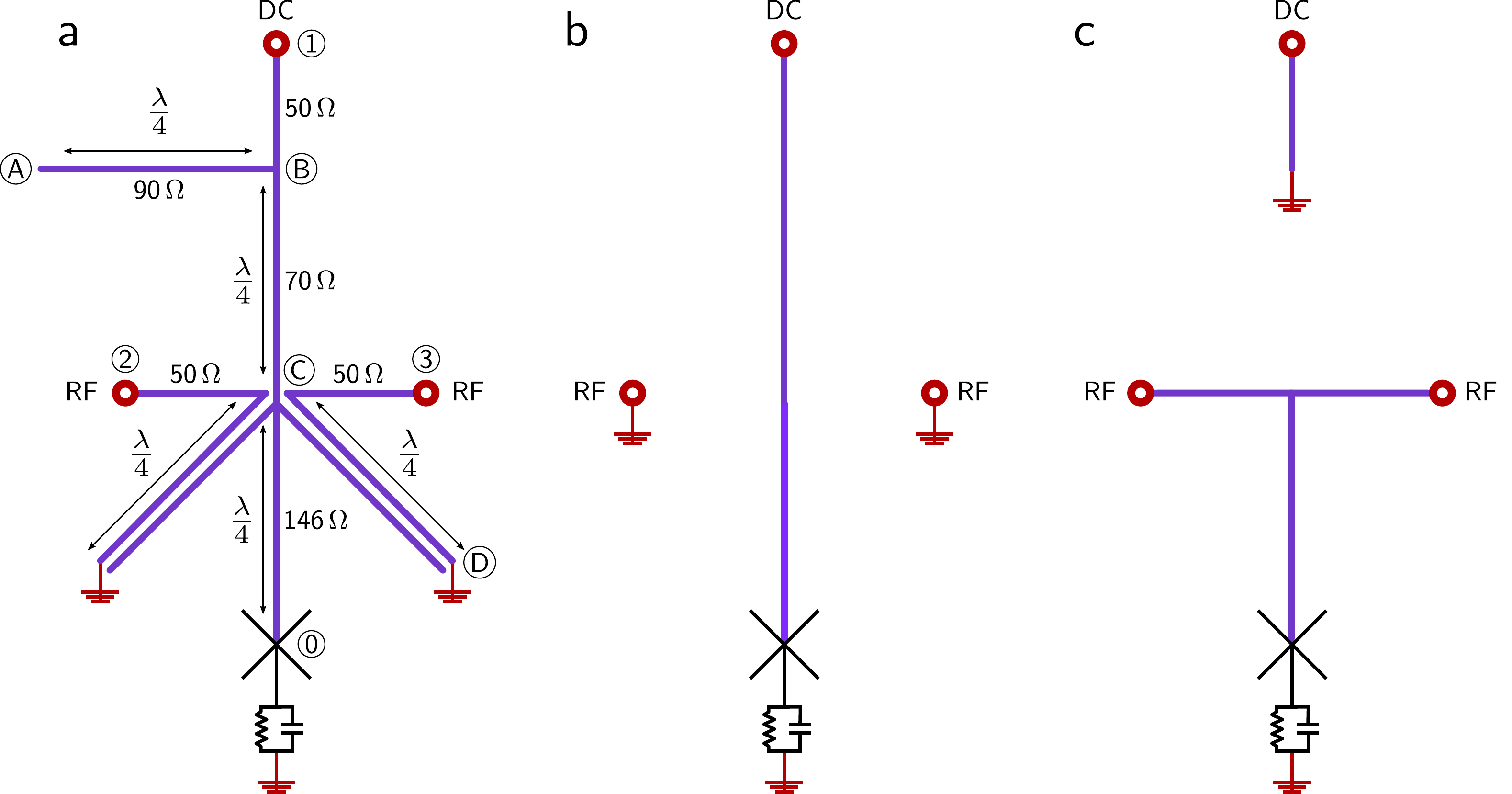}
	\caption{\textbf{a, }Circuit of the on-chip bias tee and beam splitter. Each simple purple line represents a CPW. Two parallel lines stand for coupled CPWs. Characteristic impedances, ports and $\lambda /4$ segments are indicated. The SQUID is marked by a cross and is grounded through the $RC$ circuit. \textbf{b, }Simplified circuit at low frequency. The two RF ports are grounded and the SQUID is connected to the DC bias port. \textbf{c, }Simple circuit at resonance frequency. The RF ports are dynamically connected to the SQUID and the DC port is dynamically grounded.}
	\label{fig:beamsplitter}
\end{figure}

\begin{figure}
	\centering
	\includegraphics[width=0.5\linewidth]{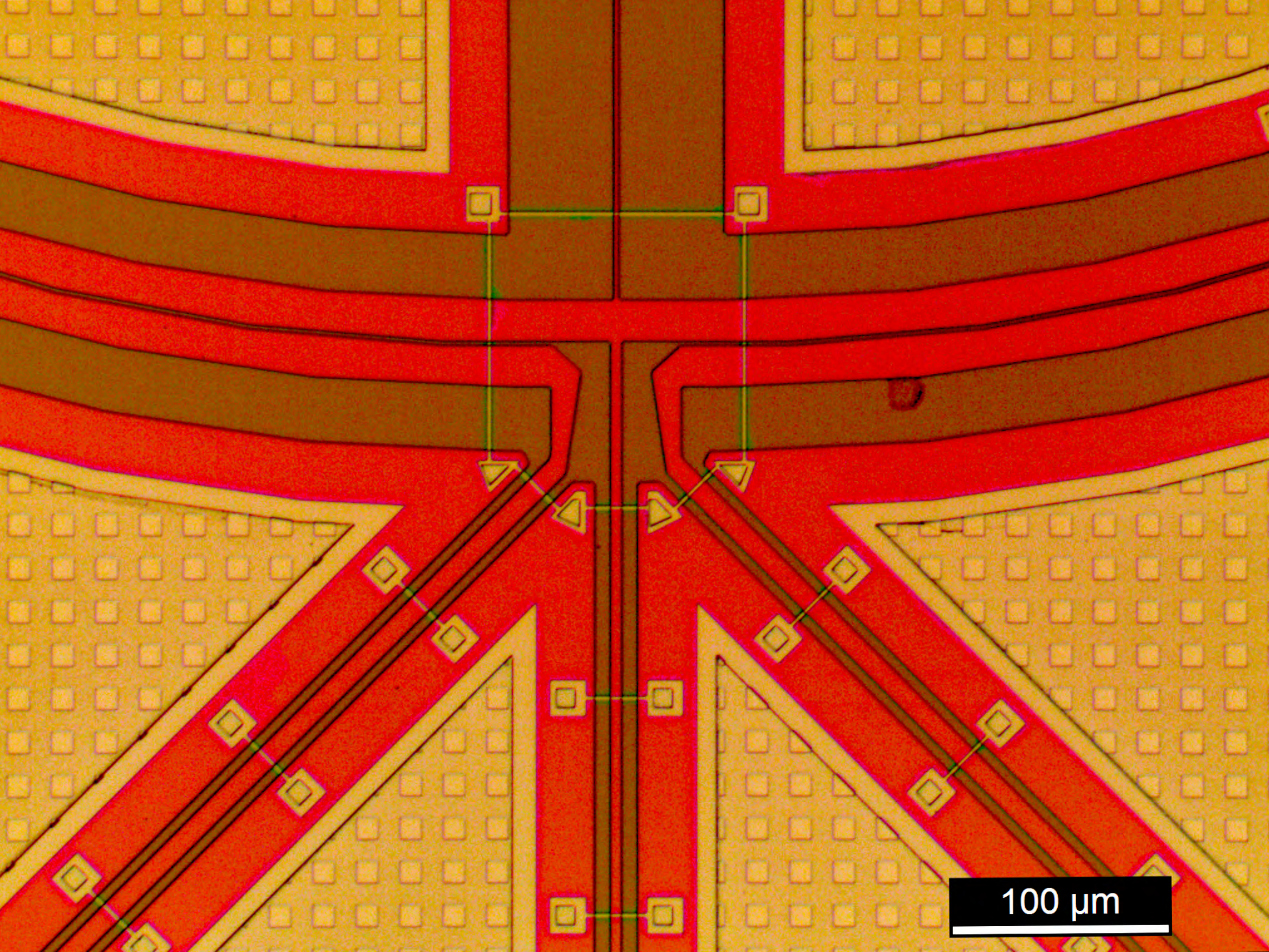}
	\caption{Optical micrograph showing the central area of the on-chip bias tee and beam splitter around point C of Fig.~\ref{fig:beamsplitter}a. CPWs are visible as red lines separated by brown gaps from the red ground planes. Ground bridges appear in yellow. The continuous yellow regions are areas where the dielectric and the top-electrode are partially etched away to avoid macro-loading effects during dry-etching. The CPWs leaving the image on the lower left and right are leading to the two RF ports. The line running straight down goes to the DC port. On the upper half of the picture the coplanar waveguides with double inner conductors are seen arching upwards. The line going straight up is the high-characteristic impedance CPW leading to the junction.}
	\label{fig:beamsplitterDetail}
\end{figure}

A key element of the device is a 4 port on-chip network of quarter-wave 
CPW segments acting as bias tee and beamsplitter. It connects the junction 
to the DC port at low frequency. At the operating frequency it connects 
it equally to two RF ports, while isolating them from each other. Figure~\ref{fig:beamsplitter} gives a schematic representation of the 
device and Fig.~\ref{fig:beamsplitterDetail} shows an optical micrograph 
of its central area. The different ports are also indicated in 
Fig.~\overview\circuit~and~\chip.

Intuitively, the working principle of the device can be understood by 
considering separately what happens at the resonance frequency $f_{0}$ 
of the $\lambda /4$ segments and at zero frequency. In the latter case, 
the only port connected to the junction is the DC bias port, while the 
RF ports are grounded (Fig.~\ref{fig:beamsplitter}b). In contrast to 
that, at $f_{0}$ the open end of the stub at A transforms to a short at 
B and to an open again at C. This means that the RF ports and the SQUID 
are isolated from the DC port and that the impedance seen by the SQUID 
will be dominated by the RF ports. At the same time, this acts as a 
filter reflecting noise leaking down the DC measurement setup.

The double lines on either side of point C are capacitively coupled 
CPWs. While at low frequency they connect the RF lines to ground and 
disconnect them from the SQUID, on resonance they act as a transformer 
connecting the SQUID at point C to a transformed RF port impedance of 
$Z_{\rm coupler} = \left(C'_{\rm C}/C_{\rm 
	T}'\right)^2\!\!Z_{\rm L} \approx \SI{24}{\ohm}$. Here $Z_{\rm L} = 
\SI{50}{\ohm}$ is the impedance of the RF ports 2 and 3, $C'_{\rm C} 
\approx \SI{160}{\pico \farad \per \meter}$ is the coupling capacitance 
between the two lines of the coupler and $C'_{\rm T} \approx 
\SI{230}{\pico \farad \per \meter}$ the total capacitance of the line 
connected to the SQUID (i.e. the sum of its capacitance to ground and to 
the other line). On resonance, the two couplers together yield a total 
impedance at point C of $Z_{\rm C} = Z_{\rm coupler}/2 = 
\SI{12}{\ohm}$. Over the whole relevant frequency range, we can roughly approximate the effective impedance generated by this bias 
tee and beamsplitter at point C by a 
$\lambda/4$ transformer with characteristic impedance of $Z_{\rm TL} = 
\sqrt{Z_{\rm C}Z_L} = \SI{24}{\ohm}$ connected to a single port of 
impedance $Z_L=\SI{50}{\ohm}$.

\begin{figure}
	\centering
	\includegraphics[width=0.5\linewidth]{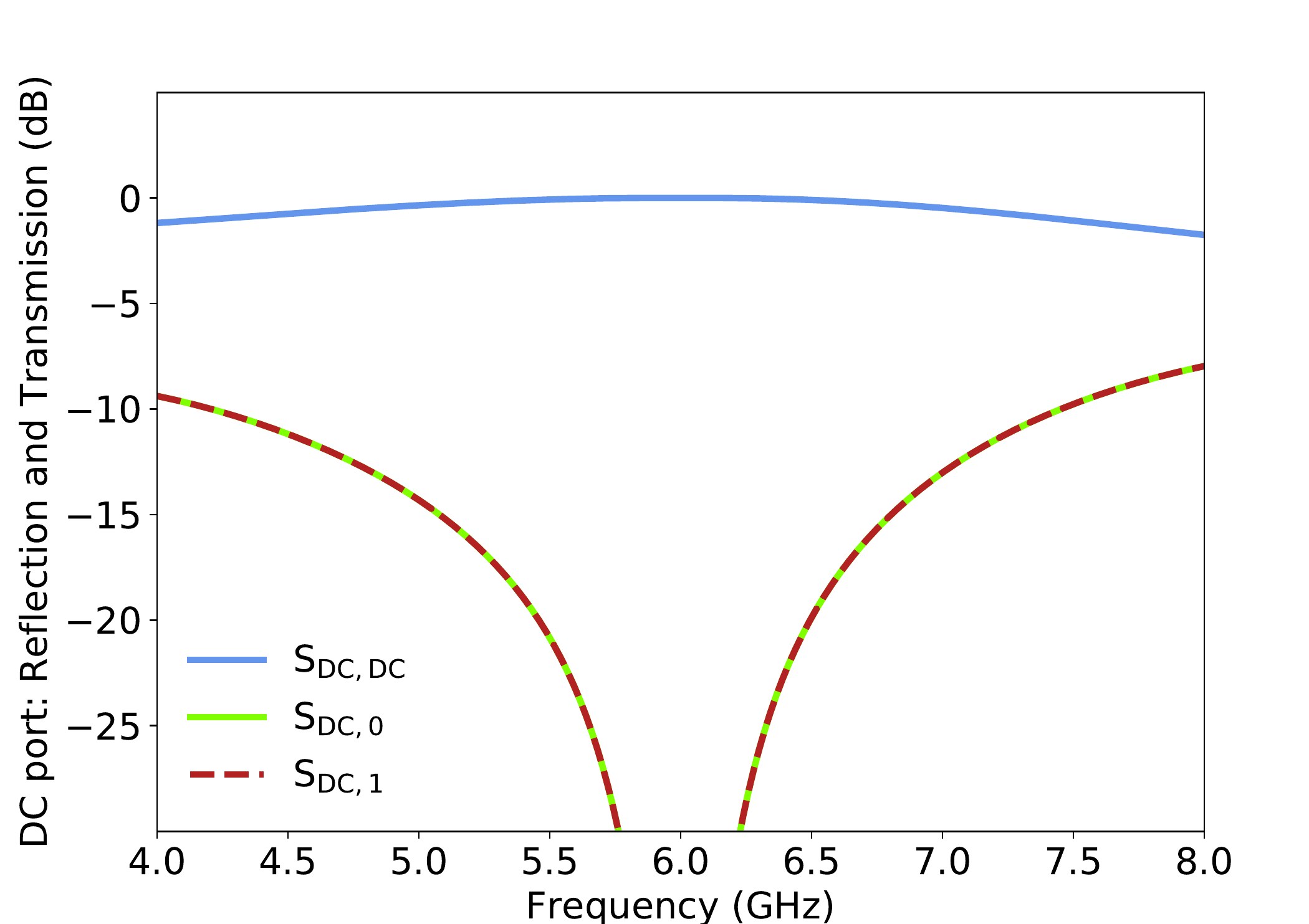}
	\caption{Isolation properties of the DC port. Solid blue line: Reflection S-parameter of the DC port. Solid green line and dashed red line: Transmission S-parameters between the RF ports and the DC port.}
	\label{fig:bsDC}
\end{figure}
\begin{figure}
	\centering
	\includegraphics[width=0.5\linewidth]{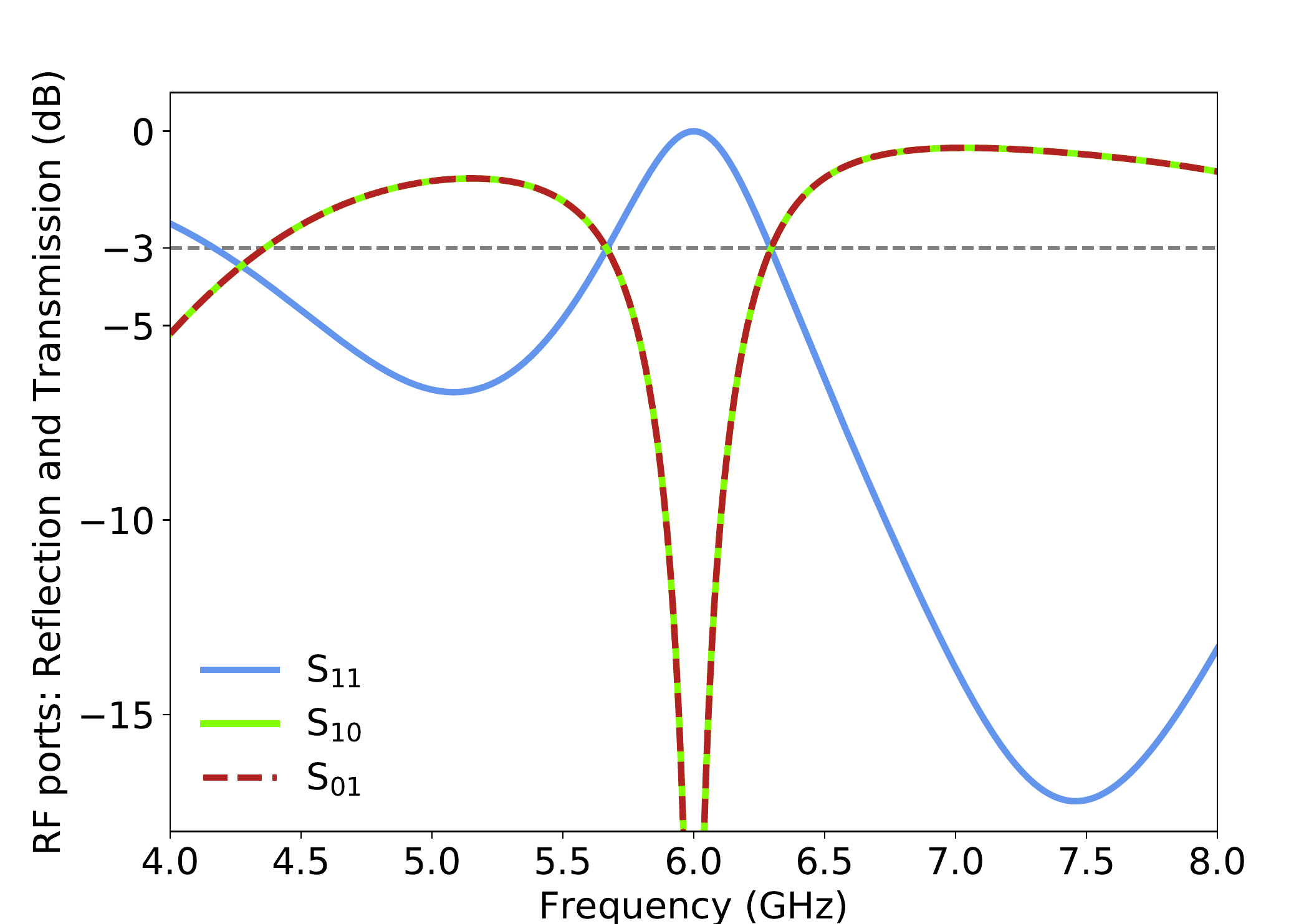}
	\caption{Isolation properties of the RF ports. Solid blue line : Reflection S-parameter of an RF port. Solid green line and dashed red line: Transmission between the two RF ports. Dashed grey line: $-\SI{3}{\decibel}$ cutoff.}
	\label{fig:bsRF}
\end{figure}
A quantitative prediction of the device parameters and frequency response was obtained using a Python-based simulation program for linear networks. The simulator has been written in our group and computes the admittance matrix of a network defined by nodes, ports and circuit elements connecting them. The latter can be either lumped elements, simple transmission lines or coupled transmission lines as the ones used in the beam splitter.

Quantities of interest are the isolation between the DC and the RF ports and the filtering of high frequency signals coming down the DC bias line (Fig.~\ref{fig:bsDC}). We perform this analysis in terms of the amplitudes of the device S-parameters using the definition $S_{ij}=20\log(|V^{-}_{i}/V^{+}_{j}|)$, where $V^{-}_{i}$ and $V^{+}_{j}$  are the incoming and outgoing complex voltage amplitudes at ports $i$ and $j$ respectively.

We first focus on the reflection of incoming signals at the DC port $S_{\rm DC, DC}$ (solid blue line) in Fig.~\ref{fig:bsDC}. It is almost unity over a wide frequency range of several $\si{\giga \hertz}$ around resonance, meaning that high frequency noise coming down the DC bias line is reflected before reaching the sample. The solid green and dotted red lines show the transmission parameters between the RF ports and the DC port. Isolation is good ($S_{DC,i}<\SI{-20}{\dB}$ over a span of $\approx\SI{1}{\giga \hertz}$) and becomes excellent around resonance.

Fig.~\ref{fig:bsRF} gives an overview of the different S-parameters concerning the RF ports. Both, the reflection of incoming signals at the ports ($S_{1,1}$, solid blue line), and the transmission between the two ports ($S_{1,0}$ and $S_{0,1}$, solid green and dotted red lines) exhibit the behaviour of a bandstop filter with a bandwidth corresponding to the full width at half maximum of the resonant structure $\approx\SI{575}{\mega \hertz}$ (see also Fig.~\psd\impedancepeak).

\subsection{Heating and cooling effects in the resistor}
\label{sec:heating}
The resistor connects to superconducting leads on both sides (SQUID and ground). Therefore electrons cannot be cooled via the leads, but the resistor is deposited directly on the MgO buffer layer in order to facilitate thermalization. Electron phonon coupling in the thick pads between the resistive lines helps to cool the electron temperature of the resistor below the smallest energy scale in the system, which is the charging energy of the capacitor ($E\sub{C}/k\sub{B}=2e^{2}/(k\sub{B}C)\approx\SI{70}{\milli \kelvin}$). This approach is based on the following assumptions:
\begin{enumerate}
	\item
	The Joule heating in the resistor can be estimated using the average current given by the $RC$ time. For a sample with \RExtracted~and \CExtractedfig~the $RC$ time is $\approx\SI{1.6}{\nano \second}$ and the resulting current is $I=2e/RC\approx\SI{0.2}{\nano \ampere}$. This leads to a Joule heating power of $I^{2}R\approx\SI{1.2}{\femto \watt}$.\\
	We also assume the film temperature to be constant in time. This is the case, because the fluctuations in current on the timescale of $RC$ are averaged out by the heat capacity of the electrons. The electronic heat capacity of a metal of volume $V$ and conduction electron density $n$ at a temperature $T$ is~\cite{ashcroft2002physique}:
	
	\begin{equation}
	C\sub{e}=\frac{\pi^{2}}{2}k\sub{B}nV\frac{T}{T\sub{F}}
	\label{equ:electronicC}
	\end{equation}
	
	Here $k\sub{B}\approx\SI{1.38e-23}{\joule \per \kelvin}$ is the Boltzmann constant and $T\sub{F}\approx\SI{5e4}{\kelvin}$ is the Fermi temperature. For one conduction electron per atom the electron density $n$ evaluates as $n=\rho N\sub{A}/A\sub{r}$, with $N\sub{A}\approx\SI{6.02e23}{\per \mol}$ the Avogadro constant, $\rho$ the volumetric mass density and $A\sub{r}$ the relative atomic mass.\\
	For Cr, $\rho\approx\SI{7.19e3}{\kilogram \per \cubic \meter}$ and $A\sub{r}\approx\SI{52e-3}{\kilogram \per \mol}$. The volume of the resistor is dominated by the cooling pads and is $V\approx\SI{2000}{\micro \meter \cubed}$. The energy needed to heat the structure up by $\Delta T=\SI{1}{\milli \kelvin}$ is $C\sub{e}\Delta T=\SI{28.3}{\electronvolt}$. Comparing this to the Joule heating power yields a rise time: $\Delta t=C\sub{e}\Delta T/P\approx\SI{4.5}{\milli \second}$. Fluctuations on the order of \si{\nano \second} can, therefore, safely be neglected.
	
	\item
	The second assumption is that the Kapitza thermal resistance originating from the coupling of the phonon populations across the boundaries between the different layers of our substrate (Si(\SI{500}{\micro \meter}):$\textrm{SiO}_{2}$(\SI{500}{\nano \meter}):MgO(\SI{20}{\nano \meter})) and the chromium pads is negligible. Following the reasoning of~\cite{Wellstood1994} we consider that a thin film cannot have an independent phonon population if its thickness is inferior to the wavelength of the most energetic phonons at a given temperature. Assuming that the Si base of the substrate is well thermalized with the copper sample holder, the combined thickness of the remaining films is $d=\SI{620}{\nano \meter}$. For $T\ll T\sub{D}=\SI{460}{\kelvin}$, where $T\sub{D}$ is the Debye temperature in Cr~\cite{ashcroft2002physique}, only acoustic phonons are relevant, which have an energy $E\sub{ph}=h v\sub{c}/\lambda=k\sub{B}T$. Here $v\sub{c}=\SI{5.9e3}{\meter \per \second}$ is the speed of sound in chromium~\cite{samsonov1968handbook}. We make the above condition more stringent by requiring that even quarter-wave resonances can be excluded. The minimum temperature needed to excite phonons of $\lambda/4 = d$ is $\approx\SI{114}{\milli \kelvin}$, a value clearly larger than the base temperature of our dilution refrigerator (\SI{15}{\milli \kelvin}).
	
	\item
	Finally, we assume that there is no heat diffusion from the resistor into the superconducting contacts. This is justified by the large gap of NbN ($T\sub{c}\approx\SI{15}{\kelvin}$). Thermal electrons are far below the gap and the dissipated power has to be evacuated through electron-phonon coupling in the cooling pads. 
\end{enumerate}

The electron temperature of a resistor with volume $V$, taking into account Joule heating and coupling to a phonon bath of temperature $T\sub{ph}$, is given by~\cite{Wellstood1994,Huard2007}:

\begin{equation}
T_{e}=\sqrt[5]{T\sub{ph}^{5}+\frac{I^{2}R}{\Sigma V}}
\label{equ:Phononcoupling}
\end{equation}

The electron phonon coupling constant $\Sigma$ has values around \SI{2e9}{\watt \per \meter \cubed \per \kelvin \tothe{5}} for most metals~\cite{Huard2007}. To work with a lower bound we consider it to be \SI{0.2e9}{\watt \per \meter \cubed \per \kelvin \tothe{5}} in our case~\cite{Giazotto2006}. The resistor has 12 cooling pads connected by 12 thin resistive line of length \SI{20}{\micro \meter} and width \SI{0.3}{\micro \meter} each. In the absence of cooling pads and for $T\sub{ph}=\SI{15}{\milli \kelvin}$ the electron temperature of the wire would be $\approx\SI{89}{\milli \kelvin}$. Taking into account the additional volume of the pads the average electron temperature of the entire structure is $\approx\SI{20}{\milli \kelvin}$. While most of the heating occurs in the lines, most of the cooling happens in the pads.

In spite of this acceptable average $T\sub{e}$, the local electron temperature in the resistive lines could still be too high. Due to their small volume, electron-phonon coupling in the lines is negligible and they are in the hot electron interaction limit~\cite{Steinbach1996}. The temperature profile along the wire in the normalized coordinate $x$ is then given by~\cite{Huard2007}:

\begin{equation}
T\sub{e}(x)=\sqrt{T\sub{ph}^{2}+\frac{3}{\pi^{2}}x(1-x)\left(\frac{eRI}{k\sub{B}}\right)^{2}}
\label{equ:electronDiffusion}
\end{equation}

The resulting temperature curve is very flat, with a difference between the $T\sub{e}$ on the borders (electron temperature of the pads) and the maximum in the middle of the wire of less than \SI{1}{\percent}.

\section{Extraction of the correlation functions}
\subsection{Cross correlation measurements}
\label{sec:Gfunctions}
Following the argumentation presented in~\cite{daSilva2010,Eichler2012} we combine quadrature measurements from the two different measurement channels to compute the correlation functions of the field emitted by our sample while rejecting contributions arising from amplifier noise. Here, we give a brief overview of the guiding principle behind this approach using a very general noise model to show that unwanted residual noise can be subtracted during data treatment through combinations of simple ``On/Off'' measurements.

\subsubsection{Noise model of the measurement chain}
\label{ss:noisemodel}
\begin{figure}
	\centering
	\includegraphics[width=80mm]{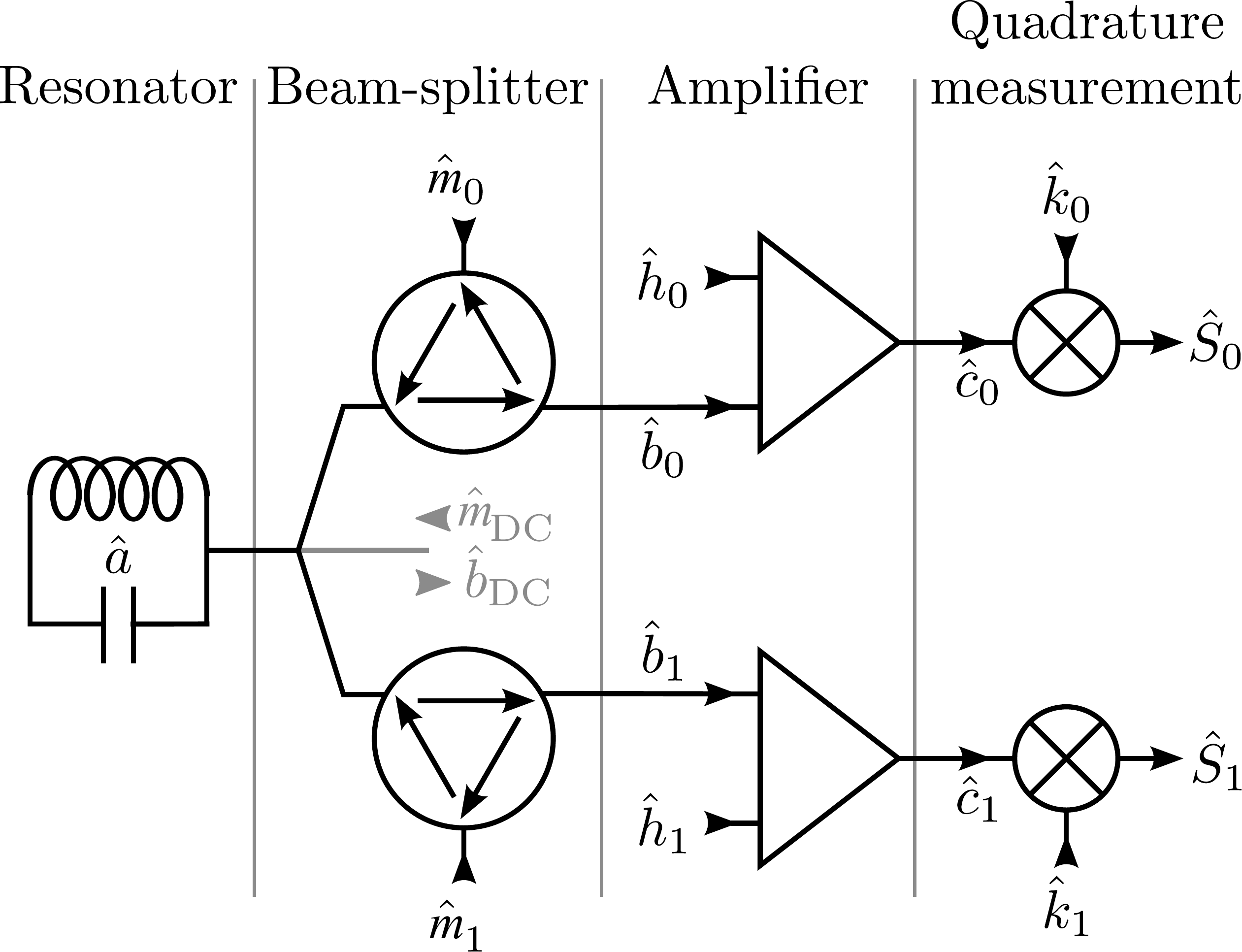}
	\caption{Schematic representation of the noise model of our measurement chain. The different symbols are explained in the text.}
	\label{fig:chain}
\end{figure}
The entire measurement process with different noise sources from the output of the sample to the quadrature measurement is summarized in Fig.~\ref{fig:chain} and contains the following steps:

\begin{enumerate}
	\item
	
	The cavity mode $a$ on our sample is connected to 3 lines (a DC line and RF lines 0 and 1). Input-output theory~\cite{walls_book} can straightforwardly be adapted to this situation by including a unitary scattering matrix $S$ describing the coupling between the ports. 
	
	\begin{equation}
	\hat{b}_i(t)=\sqrt{\gamma_i}\hat{a}(t)-\!\!\!\sum_{j = {\rm DC}, 1, 2}\!\!\! S_{ij} \hat{m}_j(t)
	\label{eq:input_output}
	\end{equation}
	
	Here $\gamma_i$ is the resonator energy decay rate into line $i = {\rm DC}, 1, 2$, $\hat{b_i}$ the outgoing field operator on line $i$ and $\hat{m}_i$ the incoming field operator. As the $S$ matrix is unitary, different output modes commute, $[b_i, b_j^\dagger] = \delta_{ij}$.

	The coupling to the DC line is designed to be negligible, $\gamma_{\rm DC} = 0$  (see Fig.~\ref{fig:bsDC}) and on resonance the two RF ports are designed to be symmetrically coupled. Therefore the cavity field leaks out into a mode
	\begin{equation}
	\hat{a}\sub{out} \approx \frac{1}{\sqrt{2}} \left(\hat{b}_1 + \hat{b}_2\right)
	\end{equation}
	with rate $\gamma \approx 2 \gamma_1 = 2 \gamma_2$. Any imbalance and residual loss can be accounted for in the gain and noise of the subsequent amplification step.

	Note that, according to \eqcite{eq:input_output}, the outgoing modes of the beam splitter are a linear combination of the incoming modes on the transmission lines which are connected to the RF ports. They are terminated by the cold loads on the isolators which are at a temperature $T\ll h\reso/k\sub{B}$ such that the input fields $\hat{m}_i$ can be considered in the vacuum state and will drop out in all normally ordered expectation values~\cite{daSilva2010} we consider below. Therefore, ideally only the sample output can produce correlations between different modes. However, we additionally fully remove any residual spurious correlation between these modes, independently of whether they arise from hot input modes or crosstalk at subsequent stages (see Sec.~\ref{ss:complexEnv}).

	\item
	In the next step the signal is amplified, which necessarily adds noise. With the power gains $g_i$ and the noise modes $\hat{h}_i$, we can write the amplified signals on either chain as~\cite{Caves1982}:
	
	\begin{equation}
	\hat{c}_i(t)  = \sqrt{g_i}\hat{b}_i(t)+\sqrt{g_i-1}\hat{h}^{\dagger}_i (t) 
	\end{equation}
	
	We take the noise on the amplifiers to be independent from their inputs giving $[\hat{b}_i,\hat{h}^{\dagger}_i]=0$ and $\langle\hat{b_i}\hat{h}^{\dagger}_i\rangle=0$. We suppose that the two noise modes commute, but we do not consider them to be uncorrelated: $\langle\hat{h}^{\dagger}_0\hat{h}_1\rangle=G^{(1)}_{\times,amp}(t,t+\tau)$. We also do not assume them to be independent from the noise of incoming modes meaning $\langle\hat{m}^{\dagger}_i\hat{h}_j\rangle\neq0$ and $[\hat{m}_i,\hat{h}_j]\neq0$, for instance due to imperfections of the isolators. No assumptions on the form of the noise cross-correlation are made.

	\item
	Lastly, quadrature measurements are performed on both outputs which introduces additional noise. One can define a complex envelope from the two quadratures, which is proportional to the input field and a noise mode~\cite{daSilva2010}:
	
	\begin{equation}
	\hat{S}_i(t) \equiv\hat{X}_i(t)+\imag\hat{P}_i(t) =\hat{c}_i(t)+\hat{k}^{\dagger}_i(t),
	\end{equation}
	
	where $\imag^2 = -1$. These complex envelope operators are defined from the classical outputs of the quadrature measurements and fulfill the equality $\langle\hat{S}_{i}(t)\rangle=\langle S_{i}(t)\rangle$, where $S_{i}(t)=X_{i}(t)+\imag P_{i}(t)$ is a complex number. The modes $\hat{k}_i(t)$ commute with the amplified fields and all the other modes except for the usual relation $[\hat{k}_i(t),\hat{k}^{\dagger}_j(t+\tau)]=\delta(\tau)\delta_{i,j}$, but they can have non-vanishing correlations.
\end{enumerate}

The noise terms intervening after the first beam splitter can be condensed into one mode $\hat{l}_i$ per channel $i$. Even though the noise from mixing will most likely be negligible compared to the amplifier noise in an experimental setup, we keep it here for the sake of completeness.

\begin{equation}
\hat{l}^{\dagger}_{i}(t)=\sqrt{\frac{g_{i}-1}{g_{i}}}\hat{h}^{\dagger}_i(t)+\frac{1}{\sqrt{g_{i}}}\hat{k}^{\dagger}_i(t)
\end{equation}

This operator still respects the commutation relations $[\hat{l}^{\dagger}_{i}(t),\hat{l}_{j}(t+\tau)]=\delta(\tau)\delta_{i,j}$ and the operators $\hat{S}_{i}$ take the form:

\begin{equation}
\hat{S}_{i}(t) =\sqrt{g_{i}}\left(\hat{b}_i(t)+\hat{l}^{\dagger}_{i}(t)\right).
\end{equation}

\subsubsection{Correlations between complex envelopes}
\label{ss:complexEnv}
The complex amplitudes contain the original field $\hat{a}$ and can be combined in different ways to obtain the correlation function $G^{(1)}$. If we were just using one measurement channel, the only option would be:

\begin{equation}
\begin{spreadlines}{8pt}
\begin{aligned}
\Gamma^{\left(1\right)}_i\left(t,t+\tau\right) &=\left<\hat{S}^{\dagger}_i\left(t\right)\hat{S}_i\left(t+\tau\right)\right> \\
&=g_i\left[\frac{1}{2}
G^{\left(1\right)}\left(t,t+\tau\right)+
G^{\left(1\right)}_{i,\textrm{noise}}\left(t,t+\tau\right)+
\delta\left(\tau\right)\right]
\end{aligned}
\end{spreadlines}
\label{equ:Gamma1dir}
\end{equation}

The last equality is derived directly from the noise model described above (see: reference~\onlinecite{Grimm2015}, Appendix~C). The first term in the last line is the desired correlation function. The second term corresponds to the summed direct correlations of all noise sources on this channel and the last term originates from the commutator $[\hat{l_i}(t),\hat{l_i}^{\dagger}(t+\tau)]=\delta(\tau)$ and can be seen as the contribution of the vacuum fluctuations of mode $\hat{l_i}$. In a realistic experiment its divergence at $\tau=0$ would be smeared out due to the necessarily finite measurement bandwidth~\cite{daSilva2010}.

A better choice to extract the correlation is:

\begin{equation}
\begin{spreadlines}{8pt}
\begin{aligned}
\Gamma^{\left(1\right)}_\times\left(t,t+\tau\right) &=\left<\hat{S}^{\dagger}_0\left(t\right)\hat{S}_1\left(t+\tau\right)\right> \\
&=\sqrt{g_0g_1}\left[\frac{1}{2}
G^{\left(1\right)}\left(t,t+\tau\right)+
G^{\left(1\right)}\sub{\times,noise}\left(t,t+\tau\right)\right]
\end{aligned}
\end{spreadlines}
\label{equ:Gamma1cross}
\end{equation}

The noise cross-correlation $G^{\left(1\right)}\sub{\times,noise}\left(t,t+\tau\right)$ is taken between the noise sources on both channels and is intuitively and experimentally (Fig.~\ref{fig:gain_noise}) much smaller than $G^{\left(1\right)}_{i,\textrm{noise}}\left(t,t+\tau\right)$. Moreover, the delta function does not emerge, since $\hat{l_0}$ and $\hat{l_1}$ commute. The extraction of $G^{\left(1\right)}\left(t,t+\tau\right)$ is a simple matter of measuring $\Gamma^{\left(1\right)}_\times\left(t,t+\tau\right)$ with the sample in the ``On'' and ``Off'' state and then subtracting the results. For the sake of completeness we have carried the quantum treatment of the field to the very end of the measurement chain. In practice, commutators of fields after the first amplification stage are negligible and a classical treatment is sufficient. Nevertheless, the noise-subtraction was performed in the case of the direct measurement, the much larger noise correlations would make the extraction very sensitive to any change in gain or amplifier noise between the ``On'' and ``Off'' measurements.

In a similar manner, one can define several combinations of complex envelopes to obtain the second order correlation function and again some include delta functions and direct noise correlations, while others do not. A choice falling into the latter category is:

\begin{equation}
\begin{spreadlines}{8pt}
\begin{aligned}
\Gamma^{\left(2\right)}_\times\left(t,t+\tau\right)
&=\left<\hat{S}^{\dagger}_0\left(t\right)
\hat{S}^{\dagger}_0\left(t+\tau\right)
\hat{S}_1\left(t+\tau\right)
\hat{S}_1\left(t\right)\right> \\
&=\frac{g_0g_1}{2}
\left[\frac{1}{2}G^{\left(2\right)}\left(t,t+\tau\right) +G^{(1)}(t,t+\tau)G^{(1)}\sub{\times,noise}(t,t+\tau) \right. \\
&\left. +G^{\left(1\right)}\left(t+\tau,t+\tau\right)G^{\left(1\right)}\sub{\times,noise}\left(t,t\right)	+G^{\left(1\right)}\left(t,t\right)G^{\left(1\right)}\sub{\times,noise}\left(t+\tau,t+\tau\right) \right.	\\
&\left. +G^{\left(1\right)}\left(t+\tau,t\right)G^{\left(1\right)}\sub{\times,noise}\left(t+\tau,t\right)
+G^{\left(2\right)}\sub{\times,noise}\left(t,t+\tau\right)
\right]	
\end{aligned}
\end{spreadlines}
\label{equ:Gamma2cross}
\end{equation}

Again, the noise cross-correlation $G^{\left(2\right)}\sub{\times,noise}\left(t,t+\tau\right)$ is the only term that remains when the sample is not active. All the other noise terms are known from the determination of $G^{\left(1\right)}\left(t,t+\tau\right)$ and can be subtracted to find the correlation $G^{\left(2\right)}\left(t,t+\tau\right)$. This measurement subtracts any system noise or spurious correlation and contains only one calibration constant, the normalization factor $\frac{g_0 g_1}{2}$. In the normalized second order correlation function $g^{(2)}$ this factor cancels as well, so that $g^{(2)}$ is fully self-calibrated without the need to know the gain and noise temperature of the measurement chain.

\section{Measurement setup}
\label{sec:msmt}
\begin{figure}
	\centering
	\includegraphics[width=0.7\linewidth]{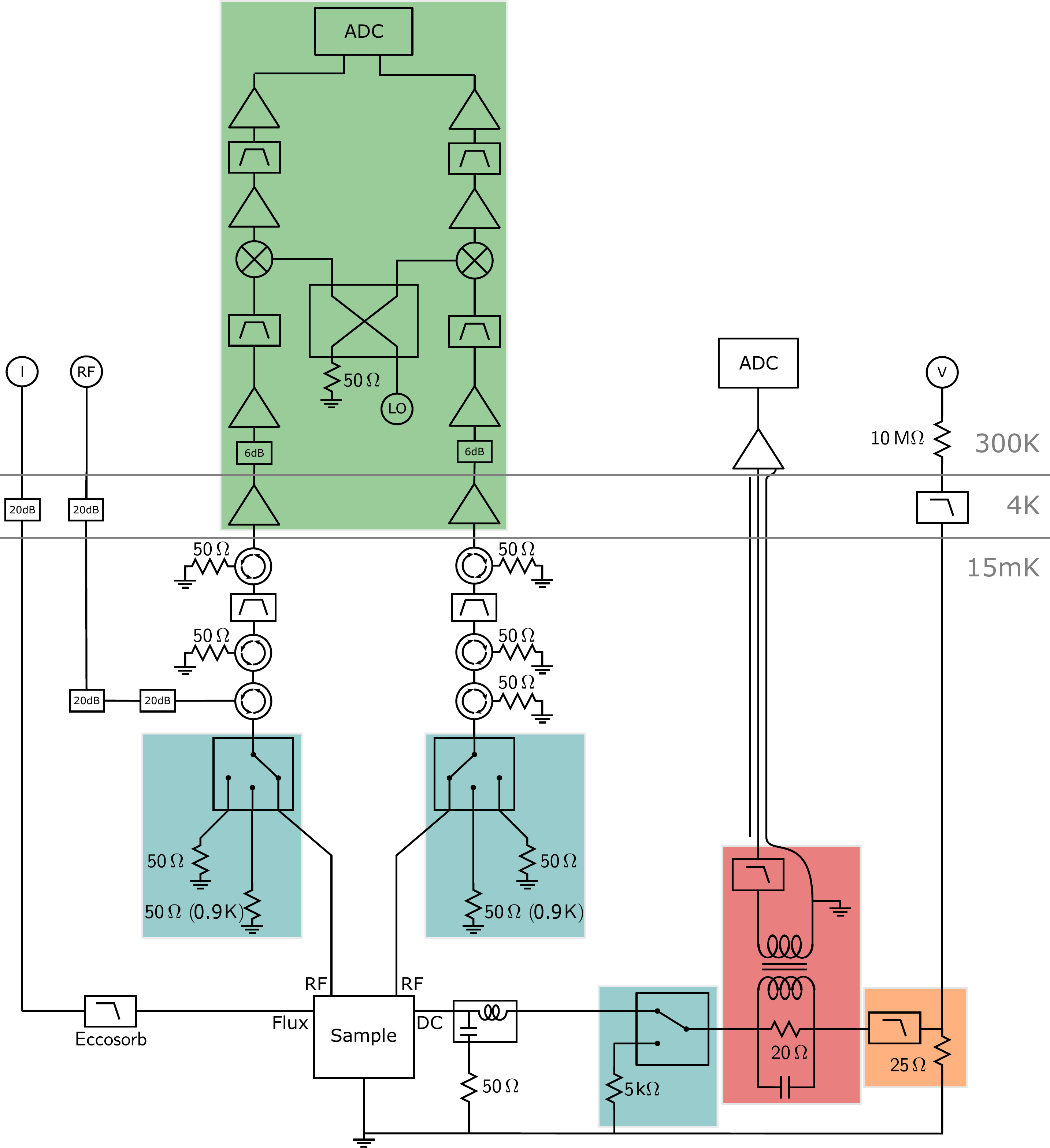}
	\caption{The entire measurement chain. Only elements inside the dilution refrigerator are represented in detail. The box labelled ``sample'' contains the device shown and described in Fig.~\overview~of the main text with corresponding ports (RF, DC and Flux). RF (green) and DC (red) measurement circuits are shown schematically and are described in the text. Blue regions indicate elements relevant to RF and DC calibration. The values of resistive elements on the lowest temperature stage are given. Unless indicated otherwise they are thermalized to the base-temperature of \SI{15}{\milli \kelvin}.}
	\label{fig:fullsetup}
\end{figure}

A schematic representation of the entire measurement chain is given in Fig.~\ref{fig:fullsetup}. Only elements inside the dilution refrigerator are described in detail. On the right side, the temperatures of the different stages are marked. The setup can be divided into the following parts:

\begin{description}
	\item[Sample] The sample under test as described in the previous sections, with two RF outputs and two inputs for the DC voltage bias and the flux (current) bias.
	
	\item[RF side] The high frequency branch of the chain includes everything beyond the two RF ports.
	\begin{description}
		\item[Calibration] 
		The blue rectangle highlights a Radiall (R591-763-600) six way switch on each channel. Its output is either connected to the sample or to other terminations used for calibration. More detail is given in section~\ref{ss:RFcal}. 
		\item[Filtering and Isolation] 
		Next on each line are bandpass filters (Microtronics BPC50403, passband between \SI{4}{\giga \hertz} and \SI{8}{\giga \hertz}) and three circulators (Raditek RADC-4.0-8.0-Cryo-S21-qWR-M2-b) used as isolators by putting a thermalized \SI{50}{\ohm} load on the third port. The lowest circulator on the left side is instead connected to an attenuated RF input line. This source is usually not active during the experiment and is only used to characterize the sample.
		\item[Amplification and measurement] 
		The remaining green section of the chain in Fig.~\ref{fig:fullsetup} starts at the cold amplifiers mounted on the \SI{4}{\kelvin} stage and ends at the analog to digital converter (ADC) of our measurement computer. It is described in section~\ref{sec:RFchain}.
	\end{description}
	\item[DC side]\hspace{\fill}
	
	\begin{description}
		\item[Diplexer]
		A Marki DPXN-M50 diplexer connects the sample DC port to the rest of the voltage bias setup at frequencies below its cross-over frequency (\SI{50}{\mega \hertz}) and to a cold \SI{50}{\ohm} termination above.
		\item[Transformer and amplifier] 
		The current passing through the sample was measured via the voltage drop it caused over a \SI{20}{\ohm} cold resistor. A CMR-direct low temperature transformer (LTT-h) with a winding ratio of $30:1$ was used to amplify the voltage across the measurement resistor and up-convert its impedance for better matching with the input impedance of the Celians (EPC1-B) amplifier. The latter has a variable gain (\SI{40}{\dB}, \SI{60}{\dB} or \SI{80}{\dB}) and a voltage noise of $\approx\SI{0.7}{\nano \volt \per \sqrthz}$. It is connected to the transformer by a Thermocoax cable. The shield of the cable is connected to ground on the base-stage and to the inverted input of the amplifier at \SI{300}{\kelvin} to avoid ground loops. The amplifier is followed by a Measurement Computing (USB-1608GX-2AO) analog to digital converter, with a maximum sampling frequency of \SI{500}{\kilo \hertz}.\\
		The transformer is housed inside a homemade filter-box including several stages of \SI{10}{\kilo \ohm} resistors, \SI{100}{\pico \farad} feed-through capacitances and homemade common mode choke. The entire system acts as a bandpass filter between \SI{300}{\hertz} and \SI{11}{\kilo \hertz} and is optimized for lock-in measurements at a frequency of $\approx\SI{1}{\kilo \hertz}$. All resistive elements are NiCr resistors (Sumusu RR1220P-XXX-D) and were tested at \SI{4}{\kelvin} displaying a maximal change in their values of $<\SI{2}{\percent}$. This element was only present during the measurement of the resistance of the $RC$ circuit.
		\item[Calibration] 
		The blue rectangle highlights a two position switch (Radiall R 572 F33 000) allowing us to connect the DC measurement to a \SI{5}{\kilo \ohm} resistor (Susumu RR1220P, change at \SI{4}{\kelvin} $<\SI{1}{\percent}$) to ground in order to calibrate the current measurement. Like the preceding component it was not present during most measurements.
		\item[Bias box] 
		The orange square represents the base temperature part of a filtered voltage divider housed in a thermalized copper box. Its input is grounded through two parallel \SI{50}{\ohm} resistors, followed by a filtering stage (\SI{4}{\micro \farad} capacitor and silver epoxy filter).
		\item[4\,K filter]
		Finally, the line is filtered at \SI{4}{\kelvin} using a home-made element containing three series \SI{10}{\kilo \ohm} resistors, with two \SI{1}{\nano \farad} capacitances to ground between them.
	\end{description}
	\item[Flux bias line]
	The flux bias line is visible at the very left side of the circuit diagram. It consists of semi-rigid cupronickel coaxial cables (Coax Co., Ltd SC-219/50-CN-CN). It is attenuated (\SI{20}{\dB}) at \SI{4}{\kelvin} and filtered at base temperature (home-made \SI{50}{\ohm}-matched Eccosorb CRS 124 low-pass filter~\cite{Slichter2009,Grimm2015}; cut-off frequency $\approx \SI{1}{\giga \hertz}$).
\end{description}

\subsection{RF measurement setup}
\label{sec:RFchain}
The main part of the high frequency measurement chain after the calibration stage and the isolators is indicated by the green box in Fig.~\ref{fig:fullsetup}. In order to be able to reject amplifier noise it has two independent channels \cite{daSilva2010, Bozyigit2010, Grimm2015} starting with the first amplification stage consisting of two Low Noise Factory cryogenic amplifiers (LNF-LNC4-8A) at the \SI{4}{\kelvin}. They work in the band between \SI{4}{\giga \hertz} and \SI{8}{\giga \hertz} and their gain and noise temperature are \SI{44}{\dB} and \SI{2}{\kelvin} respectively. The amplifiers are followed by \SI{3}{\decibel} attenuators for protection, thermalization and reduction of standing waves. The lines leading up to room temperature from the cold amplifiers are semi-rigid cupronickel coaxial cables (Coax Co., Ltd SC-219/50-CN-CN), while the lines between the base-stage and the amplifiers are niobium titanium (SC-219/50-NbTi-NbTi).

The first room temperature amplification stage uses Miteq (AMF-5F-04000800-07-10P) amplifiers with a gain of \SI{50}{\decibel} and a noise temperature of approximately \SI{50}{\kelvin}.

After some additional filtering (Microtronics BPI 17594, bandpass between \SI{4.25}{\giga \hertz} and \SI{7.75}{\giga \hertz}) the signal is down-converted using Marki (MM1-0312SS) mixers and a Rohde Schwarz SMF100A high frequency source as local oscillator, which is split on a AA-MCS power divider (AAMCS-PWD-2W-2G-18G-10W-Sf) to simultaneously act on both channels.

Then, the down-converted signal is further amplified (1 Minicircuits ZRL-2150, gain: \SI{25}{\dB} and 3 times Minicircuits ZX60-V62+, gain: \SI{15}{\dB}) and filtered (Microtronics custom filter \SI{1}{\giga \hertz}-\SI{2}{\giga \hertz}) before being digitized on an AlazarTech ATS9373 digitizer with two channels (\SI{2}{\giga \sample \per \second} and \SI{12}{\bit}).

\subsubsection{Frequency down-conversion scheme}
\begin{figure}
	\centering
	\includegraphics[width=1.0\linewidth]{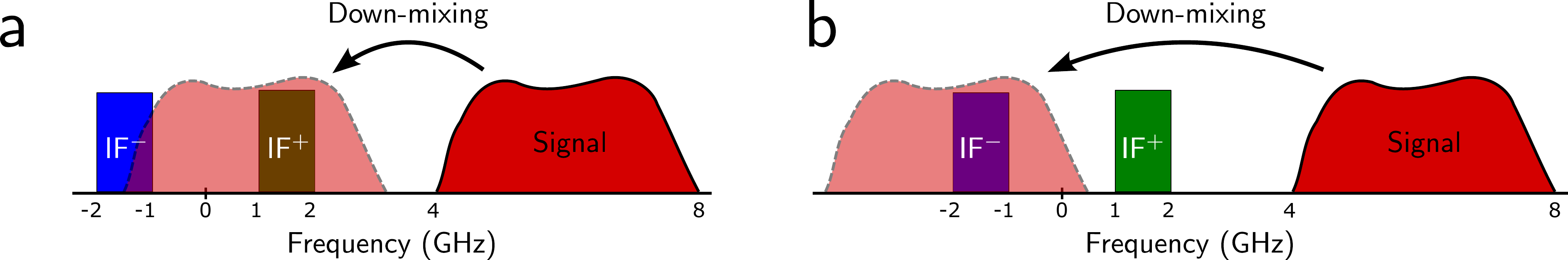}
	\caption{Down-mixing of the signal. \textbf{a, }An illustration of the case when the local oscillator (LO) frequency is chosen badly and parts of the signal are mixed simultaneously into the positive ($\mathsf{IF^{+}}$) and negative ($\mathsf{IF^{-}}$) Nyquist bands. \textbf{b, }An example of mixing only into the negative band with an LO frequency of $\approx\SI{7.5}{\giga \hertz}$.}
	\label{fig:downmixing}
\end{figure}

The correlation functions measurement presented in Sec.~\ref{sec:Gfunctions} relies crucially on having access to the time-resolved measurement record of the sample response in a bandwidth containing the entire resonator. Here we explain our particular sampling process. The signal is initially contained in a frequency band between approximately \SI{4}{\giga \hertz} and \SI{8}{\giga \hertz}, corresponding to the amplifier bandwidth. The digitizer sampling rates allow to sample slices with instantaneous bandwidth of up to \SI{1}{\giga \hertz}, given by the Shannon-Nyquist sampling theorem. This can be done in the first Nyquist band going from \SI{0}{\hertz} to \SI{1}{\giga \hertz} or in the second Nyquist band from \SI{1}{\giga \hertz} to \SI{2}{\giga \hertz}. In both cases, the other band has to be filtered out to avoid aliasing.
Figure~\ref{fig:downmixing}a shows an example of a situation where both positive and negative bands contribute to the measurement. Here the signal (shown in red) is down-converted to the position on the frequency axis indicated by the light red shape. Parts of it lie in either band and, since the ADC does not distinguish between positive and negative frequencies, are simply summed up in the final measurement result. This effect renders data extraction difficult and doubles the amplifier noise.

Figure~\ref{fig:downmixing}b shows a much more favourable situation. The signal is mixed down with an LO frequency of about $\SI{7.5}{\giga \hertz}$ and doesn't fall into the positive Nyquist band anymore. The sampled slice corresponds directly to the original signal between \SI{5.5}{\giga \hertz} and \SI{6.5}{\giga \hertz}, which is centered on the emission frequency the device discussed in this work. In a similar way we can use different local oscillator (LO) frequencies to directly and unambiguously sample the entire bandwidth of our cold amplifiers. Note that this would not be possible, if we were using the first Nyquist band. Then, the effect shown in Fig.~\ref{fig:downmixing}a could not be avoided for some parts of the signal range.

To obtain a complete map of the PSD in the \SIrange{4}{8}{\giga \hertz} range we use the following set of LO frequencies: \SIlist{3.3;3.9;4.2;4.5;7.5;7.8;8.1;8.7;9.0}{\giga \hertz}.

\subsubsection{Calibration}
\label{ss:RFcal}
Even though all correlation function measurements are self calibrated because of the normalization and noise subtraction discussed in section~\ref{sec:Gfunctions}, we still have to calibrate the power spectral density emitted by our sample in units of photons. To do so we need to know the gain and noise temperature of our output lines.

The calibration part of the RF setup is indicated by the two blue boxes on the high frequency branch in Fig.~\ref{fig:fullsetup}. On each channel it consists of a 6-port switch (Radiall R591763600) connecting the input of the amplifiers to either the sample or the two calibration references. They are two \SI{50}{\ohm} loads, thermally isolated from the switch by short NbTi coax lines and thermalized at base temperature and at the still respectively.

The noise power spectral density coming from a \SI{50}{\ohm} load on a matched line at temperature $T$ is:
\begin{equation}
S^{\textrm{(in)}}(T)=\frac{h\freq}{2}\coth\left(\frac{h\freq}{2k\sub{B}T}\right).
\end{equation}
The total signal on channel $i$ after amplification is:
\begin{equation}
S^{\textrm{(out)}}_i=g_i\left(S^{\textrm{(in)}}(T_i)+N_i\right)
\label{equ:noiseT}
\end{equation}

Here, $g_i$ and $N_i$ are the amplifier gain and the combined effective noise photon number of the amplifier and the cables leading up to it from the \SI{50}{\ohm} load.

At base temperature ($T\sub{base}\approx\SI{15}{\milli \kelvin}$), $S^{\textrm{(in)}}(T) \approx \frac{1}{2}h\freq$, i.e.~approximately independent of temperature, so that accurate knowledge of the temperature of the cold loads is not essential. At the still temperature ($T\sub{still}\approx\SI{0.9}{\kelvin}$), however, the emitted noise is $S^{\textrm{(in)}}(T)\approx k\sub{B}T$. Therefore, we track the temperature of the loads at still temperature with a dedicated germanium thermometer mounted on their thermalisation copper bracket.

By measuring the signal when switched to the still load ($S\sub{still}^{\textrm{(out)}}$) as well as the signal coming from the base stage load ($S\sub{base}^{\textrm{(out)}}$), we can extract both gain and noise photon number of each measurement chain:
\begin{equation}
g=\frac{S\sub{still}^{\textrm{(out)}}-S\sub{base}^{\textrm{(out)}}}{S^{\textrm{(in)}}(T\sub{still})-S^{\textrm{(in)}}(T\sub{base})}
\end{equation}
\begin{equation}
N=\frac{S\sub{base}^{\textrm{(out)}}}{g} - S^{\textrm{(in)}}(T\sub{base})
\end{equation}

\begin{figure}
	\includegraphics[width=\linewidth]{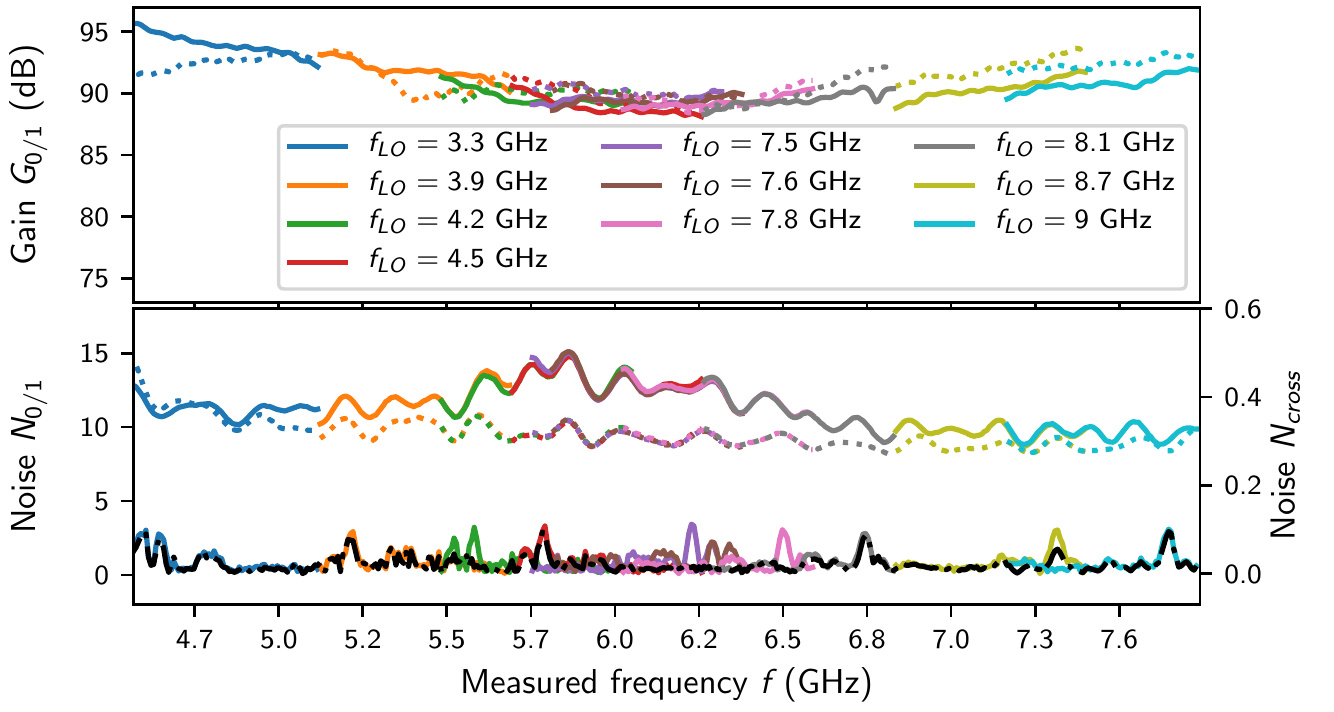}
	\caption{Calibration of the amplification chain. Measuring the power spectral density at the output
		of our amplification chain for 2 different noise references (see text) allows us to determine the gain and noise in units of photons of the
		chain for each frequency. Different colours represent measurements performed for different local oscillators used in the experiment. The two different measurement channels are shown as solid and dotted curves. 
		The lower curves in the bottom panel indicate cross noise between channels when the they are connected to the load at base temperature (coloured curves) or connected to the sample (black curve).}  
	\label{fig:gain_noise}
\end{figure}

Since our measurement is spectrally resolved, this can be done for each frequency point and of course for each local oscillator on both channels, thus fully calibrating the system.

Figure~\ref{fig:gain_noise} shows the result of a typical gain calibration. Different colours represent different local oscillators used for the measurement shown in the main text.  The lowest coloured curves in the bottom panel show the cross noise photon number between both channels, which was obtained by combining complex envelopes from both channels similar to the correlation function measurements described above. It is 0 within the uncertainty of our measurement. The black dashed-dotted curve on top of them shows the cross noise measured when the sample with no applied voltage bias (in the ``off'' mode) served as the low temperature reference instead of the cold \SI{50}{\ohm} loads.

\subsubsection{Drift compensation for on-off measurements}
This calibration scheme involves commuting the switches fixed to the base stage of our dilution refrigerator and cannot be repeated too often to avoid heating. In between calibrations, the gain of the amplification chain and phase delay may drift, mainly due to temperature fluctuations. To compensate for these drifts we use the fact that in all measurements (PSD, $g^{(1)}$ and $g^{(2)}$) presented in this article we periodically turn the bias voltage on and off, with a period of the order of 1 second. We do so in order to be able to remove all unwanted cross-correlation terms not due to the signal emitted by the sample. In the ``off'' part of the cycle when the sample does not emit photons, we continuously measure the noise of each amplification chain. This noise is dominated by the HEMT noise which is very insensitive to the actual physical temperature of the amplifier and therefore provides a more stable reference than the gain of the full chain containing many active components. We therefore renormalize the gain as follows:

\begin{equation}
g\sub{inst}=\frac{S^{\rm (out)}\sub{off}}{N}
\label{equ:Ginst}
\end{equation}
In this expression $S^{\rm (out)}\sub{off}$ is the measured signal when the sample is switched off, averaged over several cycles. Thus, we can account for slow drifts in the gain of the entire chain as illustrated in the left panel of Fig.~\ref{fig:g1phase} for a measurement lasting several days.

\subsubsection{Additional drift compensation for $G^{(2)}$ measurements}

\begin{figure}
	\includegraphics[width=0.7\linewidth]{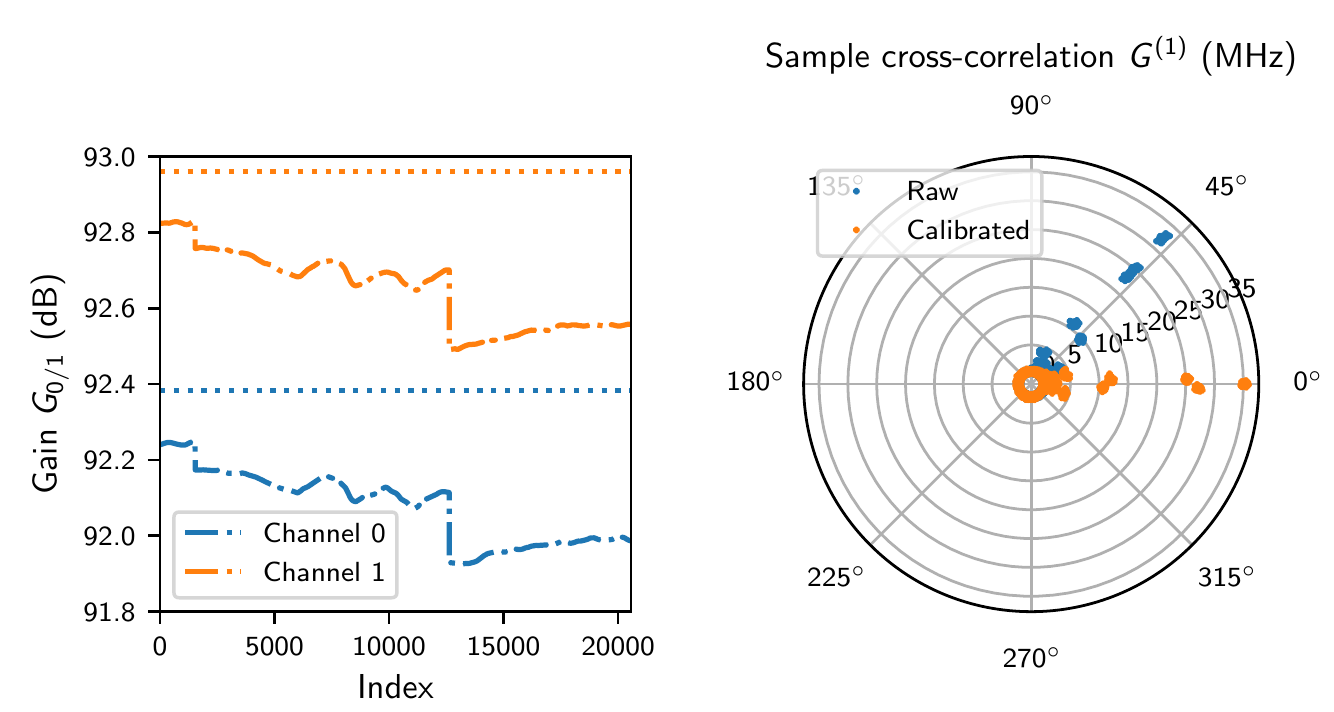}
	\caption{Gain and phase-drift compensation. Left panel: the dotted lines show the gain based on initial calibration. The dash-dotted lines are instantaneous gains according to \eqcite{equ:Ginst} measured over several days. Abrupt jumps at index 1000 and 12000 correspond to interruptions in acquisition for several hours or days. Right panel: Polar coordinate plot of several complex valued $G^{(1)}(\tau)$ measurements with each group of points representing a specific time delay $\tau$. The blue dots show the averaged original data. After compensation of gain and phase drifts (see text) data points align on the real axis (orange dots).}  
	\label{fig:g1phase}
\end{figure}

The correlation function measurements performed according to Eqs.~(\ref{equ:Gamma1cross}) and (\ref{equ:Gamma2cross}) are complex valued with a trivial complex phase given by the difference in phase of the amplitude gain of the two amplification chains. These measurements often last several days and in order to average them correctly, any phase drift must be removed prior to averaging.

To do so, we use the fact that the $G^{(1)}$ measurement has several orders of magnitude lower statistical noise than the $G^{(2)}$ measurement, so that any changes in $G^{(1)}$ averaged over several minutes are due to slow drifts. Fig.~\ref{fig:g1phase} shows such a measurement. Each group of blue dots corresponds to a specific time delay $\tau$, with each point representing an averaged measurement. The group of points with the largest magnitude corresponds to $\tau=0$. We first compensate for magnitude drifts as described above and then shift the phase of the $G^{(1)}$ measurements by $\delta_i + 2\pi\Delta f_i t$ by fitting to the points at $\tau = 0$, $\SI{\pm 1}{\nano\second}$, $\SI{\pm 2}{\nano\second}$ for each averaged measurement, indexed by $i$. $\delta_i$ accounts for drifts in the gain phase, $\Delta f_i$ is the difference between demodulation frequency and resonator frequency. After applying these compensations for magnitude and phase drift we obtain the $G^{(1)}$ curves marked by orange points which align on the real axis.

\subsection{Numerical signal processing}
Signals are bandpass filtered between \SI{1}{\giga\hertz} and \SI{2}{\giga\hertz} and recorded by an AlazarTech ATS9373 digitizer on 2 channels at a rate of 2\,GSamples/s and a resolution of 12\,bits. This data is streamed to PC memory and analyzed in real time on 16 CPU cores working in parallel. First we subtract from each channel the data recorded on the other channel convoluted with an appropriate finite impulse response filter (16 taps) in order to reduce crosstalk occurring on the card from approximately \SI{-40}{\dBc} to approximately \SI{-60}{\dBc}.

\subsubsection{Power spectral density measurements}
Power spectral density calculations are performed by calculating fast Fourier transforms $f_{i,n}$ of the data in channel $i=0,1$ for each block of data $n$. We then calculate the power spectral densities ${\rm PSD}_{i,j} = \sum_n f^*_{i,n}f_{j,n}$.

${\rm PSD}_{\rm i,i}$ is the power spectral density of channel $i$. ${\rm PSD}_{0,1} = {\rm PSD}_{1,0}^*$ is the cross power spectral density. As all signals have low coherence and low dynamic range, we do not apply any windowing in order to get the best signal to noise ratio.

\subsubsection{Correlation function measurements}
The correlation functions $G^{(1)}$ and $G^{(2)}$ are calculated from complex amplitudes $S_i$ of the output signal of channels $i=0,1$ according to Eqs.~(\ref{equ:Gamma1cross}) and (\ref{equ:Gamma2cross}). We describe here how these envelopes are calculated and how $G^{(1)}$ and $G^{(2)}$ are computed numerically.

At the output of the amplification chain we have voltage signals $V_i(t)$ centered around the resonator frequency $f_0$. For the $g^{(1)}$ and $g^{(2)}$ we need the complex time-dependent envelope $S_i(t)$ of this signal:

\begin{equation}
V_i(t) = \frac{1}{2}\left(S_i(t) \exp^{-\imag2\pi f_0 t} + S^*(t) \exp^{\imag2\pi f_0 t}\right) 
= \Re{S_i(t)} \cos(2\pi f_0 t) - \Im{S_i(t)} \sin(2\pi f_0 t)
\end{equation}

We down-convert this signal with a local oscillator at $f_{\rm LO}$. 
\begin{eqnarray}
V_{i,\rm IF} &=& \left( V_i(t) \cos(2\pi f_{\rm LO}t) \right) * \sqcap\\
&=& \Re{S_i(t)} \left(\cos(2\pi( f_0 - f_{\rm LO}) t) + \cos(2\pi (f_0 + f_{\rm LO}) t)\right) * \sqcap\\
&-& \Im{S_i(t)} \left(\sin(2\pi( f_0 -  f_{\rm LO})t) + \sin(2\pi (f_0 + f_{\rm LO})t)\right) * \sqcap\\
&=& \left(\Re{S_i(t)} \cos(2\pi (f_0 - f_{\rm LO}) t) - \Im{S_i(t)} \sin(2\pi( f_0 - f_{\rm LO})t)\right) * \sqcap
\end{eqnarray}
Where $\sqcap$ is the bandpass filter matching the 2nd Nyquist band of our digitizer from $\frac{1}{2\Delta T}$ to $\frac{1}{\Delta T}$ with $\Delta T = \SI{500}{\pico\second}$ the sampling interval of our digitizer. We choose the local oscillator frequency such that the down converted signal is at the center of our filter, i.e. 
\begin{equation}
f_0 - f_{\rm LO} = \frac{3}{4\Delta T}.
\end{equation}

The sampled voltages $V_{i,n} = V_{\rm IF}(n\Delta T)$ are
\begin{eqnarray}
V_{i,n} &=& \left(\Re{S_i(t)} \cos(3\pi n/2) - \Im{S_i(t)} \sin(3\pi n/2)\right)* \sqcap\\
&=&\left\{\begin{array}{ll}
R_{i,n/2} (-1)^{n/2} &\quad(n\;\mathrm{even})\\
I_{(i,n-1)/2} (-1)^{(n-1)/2} &\quad(n\;\mathrm{odd})\\
\end{array}\right.
\end{eqnarray}
with $R_{i,m} = \Re S_i(2m \Delta T)$ and $I_{i,m} = \Im S_i((2m+1) \Delta T)$. 
We then apply a finite impulse response filter kernel $k = \{0.042,\;-0.338,\;0.469,\;-0.150,\;0\}$ to the $R_{i,m}$ quadrature (i.e.\ $R_{i,m}' = \sum_{l=0}^4 R_{i,m-l} k_l$) and the time-reversed kernel to the $I_{i,m}$ quadrature (i.e.\ $I_{i,m}' = \sum_{l=0}^4 I_{i,m-l} k_{4-l}$), in order to address the following issues:
\begin{enumerate}
	\item The real and imaginary part of the envelope are not measured at the same time: $\Re S_i$ at even n and $\Im S_i$ at odd n, i.e.\ they are shifted by one sampling interval $\Delta T$. The center of the kernel is at $\Delta T/2$ so that the filtered quadratures are centered at the same time. 
	\item The bandpass filter $\sqcap$ has a time response close to a sinus cardinalis, with strong side lobes which can lead to counter-intuitive long-time correlations features. The finite impulse response filter smoothes this response.
	\item The filter is slightly wider than our resonator and therefore accepts unnecessary amplifier noise, degrading signal to noise ratio and dramatically increasing data acquisition times. The finite impulse response filter closely matches the resonator bandwidth.
\end{enumerate}

By combining the quadratures we then obtain the desired complex envelopes $S_{i,m}$ for channels $i=0,1$ sampled at $\left(m+\frac{1}{2}\right) \Delta T$.

In order to calculate the $\Gamma^{(1)}_\times$ and $\Gamma^{(2)}_\times$ according to Eqs.~(\ref{equ:Gamma1cross}) and (\ref{equ:Gamma2cross}) we first calculate the fast Fourier transform of a block of $N$ data points of $S_{0,m}$, $S_{1,m}$ and $S_{0,m} S_{1,m}$ padded with an equal amount of 0 (necessary to avoid wrap-around), which we call, respectively $F_{0,n}$, $F_{1,n}$ and $F_{\times,n}$. The Fourier transform $F^{(1)}$ of $\Gamma^{(1)}_\times$ is then $F_i^{(1)} = F^*_{0,n} F_{1,n}$ and the Fourier transform $F^{(2)}_n$ of $\Gamma^{(2)}_\times$ is $F^{(2)}_n = F_{\times,-n} F_{\times,n}$. These Fourier transforms are then averaged over many blocks of size $N$. The inverse Fourier transform is performed in post processing and a scaling factor $(N-|\tau/2\Delta T|)^{-1}$ is applied in order to account for reduced overlap in the finite convolution product with increasing $\tau$.  

\section{Error analysis}

\begin{figure}
	\centering
	\includegraphics[width=0.8\linewidth]{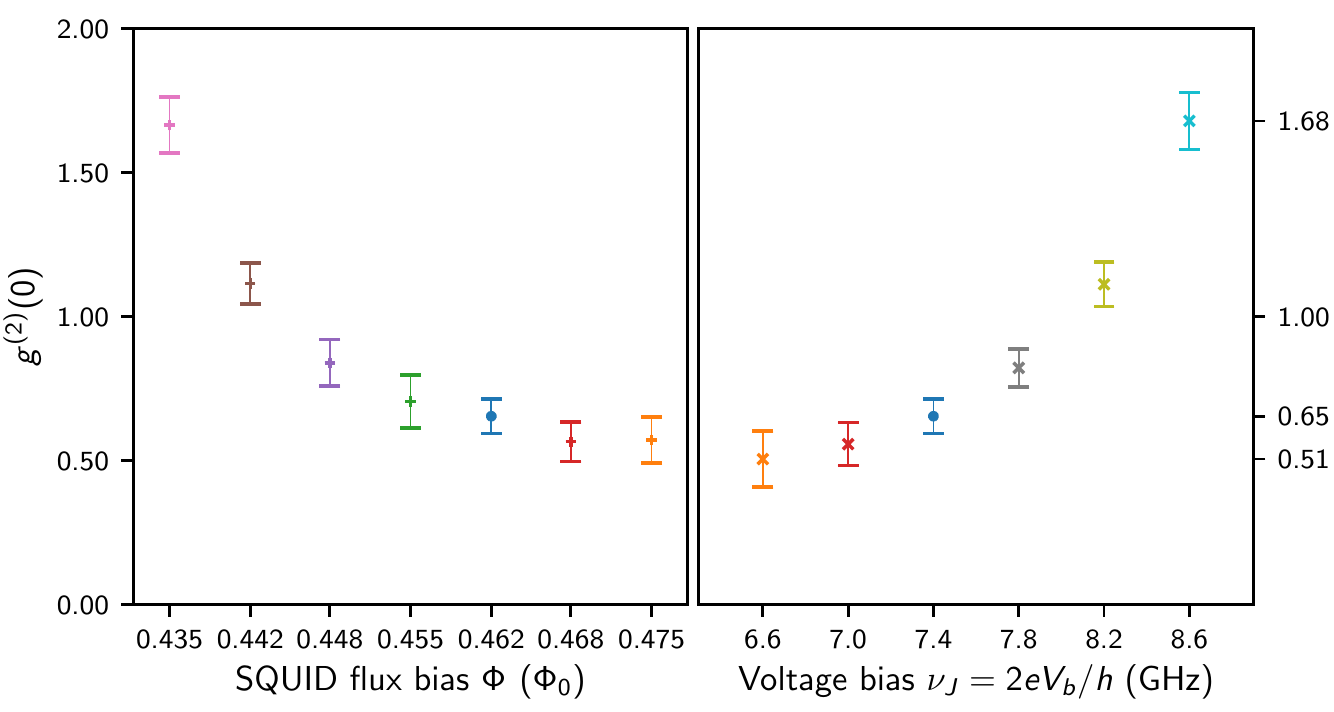}
	\caption{Statistical error of the $g^{\left(2\right)}\left(0\right)$ measurements shown in Fig.~\gfree~of the main text. Left panel: Mean values with $\pm3\sigma$ error bars of $g^{\left(2\right)}\left(0\right)$ as a function of flux bias corresponding to the data shown in Fig.~\gfree b. Right panel: The same quantities as a function of voltage bias corresponding to the data shown in Fig.~\gfree c.}
	\label{fig:error}
\end{figure}

The number of averages performed for each measurement presented in Fig.~\psd,~\gfree,~\gpulsed~ of the main text is given in table~\ref{tab:avg}. We estimated the statistical error of our data by saving averages over $n\sub{b}$ blocks of $n\sub{avg}$ samples on each channel. The standard deviation of the final result is then $\sigma=\sigma\sub{b}/\sqrt{n\sub{b}-1}$, where $\sigma\sub{b}$ is the standard deviation of the results for each block. We found the error of the first order correlation function measurement to be negligible (e.g. for $G^{\left(1\right)}\left(0,0\right)=\SI{94.82}{\mega \hertz}$ in Fig.~\gfree a, b the statistical error is $\pm3\sigma=\pm\SI{0.05}{\mega \hertz}$). This also holds for the power spectral density and emission rate data shown in Fig.~\psd~ and Fig.\gfree a. For the second order correlation function $g^{\left(2\right)}\left(0,\tau\right)$ the error is significant because the number of averages necessary for a given signal to noise ratio scales exponentially with the order of the measured correlation function (see reference~\onlinecite{daSilva2010} and reference~\onlinecite{Grimm2015} Ch.3). In Fig.~\ref{fig:error} we show $g^{\left(2\right)}\left(0\right)$ for each curve in Fig.~\gfree b and Fig.~\gfree c of the main text together with the extracted error bars ($\pm3\sigma$). The point at $\tau=0$ was specifically chosen because the measured signal is weakest and the amplifier noise correlation highest at this value making the error bars an upper bound for the other data points.
\begin{table}[h!]
	\centering
	\begin{tabular}{c|c|c|c|c|c}
		Figure 
		& $\log_{2}(n\sub{avg})$
		& $\log_{2}(n\sub{points})$
		& $n\sub{b}$ 
		& $G^{\left(1\right)}\left(0,0\right)$ (\si{\mega \hertz})
		& $3\sigma\sub{G1}$ (\si{\mega \hertz})\\
		\hline
		\psd\psdfreq  & 32 & 8 & 64 &  &  \\ \hline
		\psd\psdflux  & 32 & 8 & 1 &  &   \\ \hline
		\gfree a  & 32 & 8 & 4 &  &  \\ \hline
		\gfree b  & 32 & 7 & \makecell{36867, 26181,\\20486, 10961,\\9430,21891,\\32564} & \makecell{26.484, 31.114,\\35.198, 37.489,\\36.910, 32.489,\\25.515} & \makecell{0.005, 0.006,\\0.006, 0.005,\\0.008, 0.006,\\0.005}  \\\hline
		\gfree c  & 32 & 7 & \makecell{48405, 25304,\\20486, 14653,\\14659, 18551} & \makecell{22.158, 29.881,\\35.198, 36.669,\\33.931, 28.088} & \makecell{0.004, 0.005,\\0.006, 0.007,\\0.007, 0.007}  \\ \hline
		\gpulsed a  & 30 & 7 & 3357 & 94.82 & 0.05 \\ \hline
		\gpulsed b  & 30 & 7 & 3357 & 94.82 & 0.05 \\ \hline
		\gpulsed c  & 30 & 7 & 3357 & 94.82 & 0.05 \\
	\end{tabular}  
	\caption{Summary of measurement statistics for all relevant figures of the main text. Fourier transforms for correlation functions and PSD were performed over $n\sub{points}$ samples. For each curve in the final results, $n\sub{b}$ blocks of $n\sub{avg}$ samples were averaged. The mean value and standard deviation were calculated from these $n\sub{b}$ values. Where relevant, the mean value of $G^{\left(1\right)}\left(0,0\right)$ and three standard deviations $3\sigma\sub{G1}$ are given. Multiple values correspond to the different curves in the figure from bottom to top.}
	\label{tab:avg}
\end{table}

\section{Independent measurement of the dark region in Fig.~\psd\psdflux}
\label{sss:darkregion}
We have performed a complementary measurement of the dark region in Fig.~\psd\psdflux~of the main text by measuring the flux- and voltage bias dependent transmission through the two RF ports of the sample with a vector network analyzer (VNA). This was done by sending the stimulus to the sample via the circulator of one of the amplifier chains (visible on the left side of Fig.~\ref{fig:fullsetup}) and then measuring the amplified response on the other channel.

\begin{figure}
	\centering
	\includegraphics[width=0.6\linewidth]{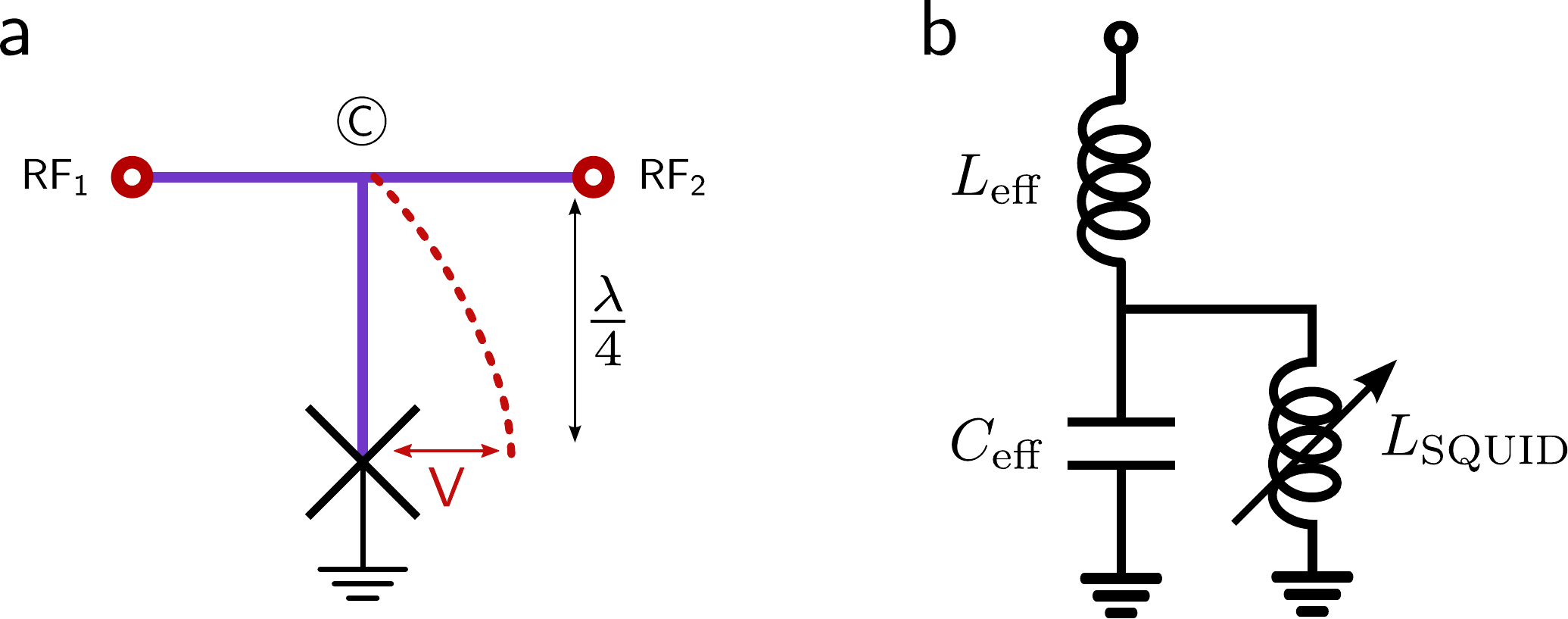}
	\caption{Model of the resonance observed in the VNA transmission measurements. \textbf{a, }Sketch of the relevant parts of the on-chip beam splitter from section~\ref{design}. Point C is the same as indicated in Fig.~\ref{fig:beamsplitter}. The $\lambda/4$ segment is labelled and the SQUID is represented by a cross. A dashed red line sketches the voltage profile on resonance. \textbf{b, }The lumped circuit element model of the $\lambda/4$-resonator and the SQUID consisting of the effective capacitance ($C\sub{eff}$) and inductance ($L\sub{eff}$) of the resonator as well as the tunable inductance of the SQUID ($L\sub{SQUID}$).}
	\label{fig:mode}
\end{figure}
\begin{figure}
	\centering
	\includegraphics[width=0.7\linewidth]{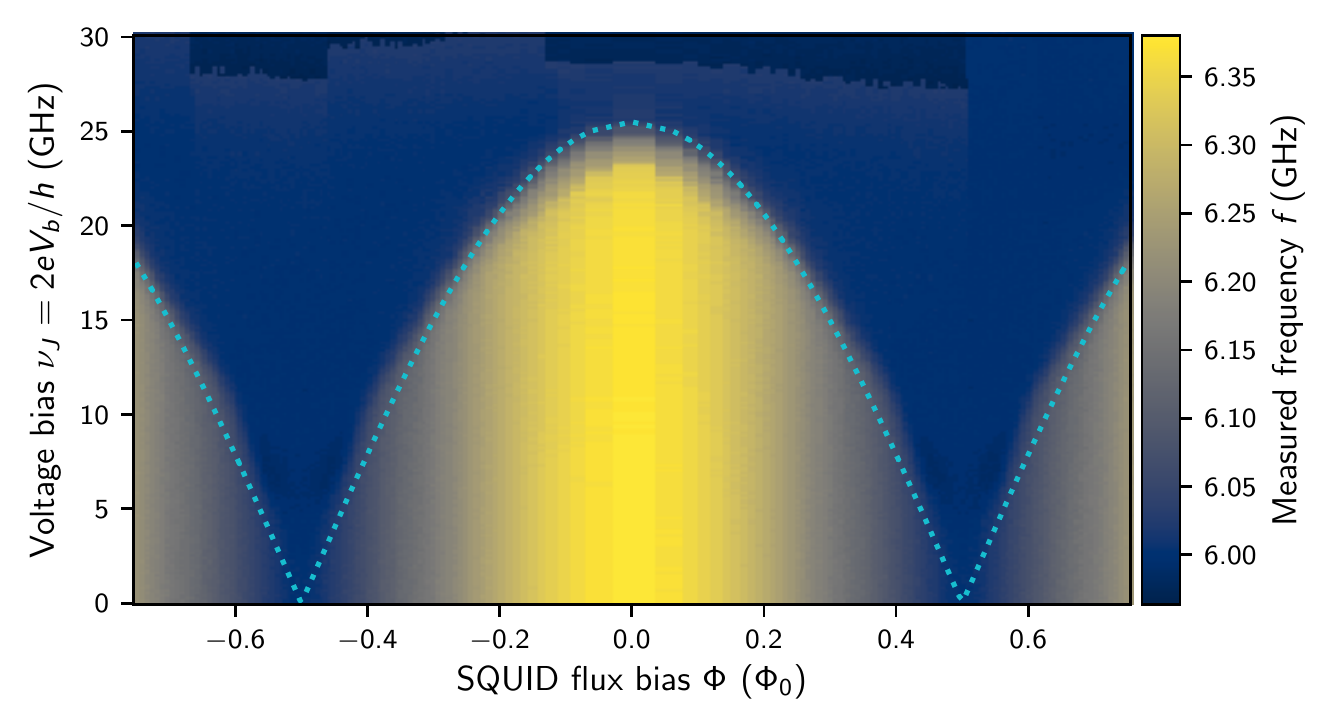}
	\caption{Frequency of the anti-resonance in a VNA transmission measurement between the two RF ports of the sample as a function of flux in units of flux quanta $\Phi_0$ and voltage bias. The frequency becomes flux-tunable where the SQUID is on its current branch and has a finite contribution to the inductance of the stub shown in Fig.~\ref{fig:mode}. The dotted curve is a function $|\cos(\pi \Phi/\Phi_{0})|$, rescaled so its apex matches the top of this region. Outside of the boundary the SQUID is on the voltage branch and the frequency of the anti-resonance does not depend on flux.}
	\label{fig:vnaFull}
\end{figure}
As discussed in section~\ref{design}, we expect the two RF ports to be decoupled on resonance, where the last $\lambda/4$ element between the node at point C and the SQUID acts as an open stub~\cite{pozar2011microwave}. Figure~\ref{fig:mode}a gives a schematic representation. It only shows the parts of the on-chip beam splitter and bias tee relevant to this measurement. The horizontal section corresponds to the two arms going down at \SI{45}{\degree} angles in Fig.~\ref{fig:beamsplitter} coupling point C to the RF ports. At the resonance frequency of the device the stub decouples the two RF ports leading to an anti-resonance dip in the transmission measurement (Fig.~\ref{fig:bsRF}).

From Fig.~\ref{fig:mode}a we can see that the SQUID is at the open end of the resonator, where the voltage is at its maximum, while the current is zero. This means that it is situated at the point of maximum impedance on resonance. This resonant circuit can be modelled by a series LC-circuit, since it produces a dip in the measured signal. The inductance coming from the SQUID shunts the effective capacitance to ground as shown in Fig.~\ref{fig:mode}b. Here, we neglect the impedance of the $RC$ circuit in series with the SQUID because we suppose the junction impedance to dominate. The anti-resonance in the measurement occurs at the frequency where the total impedance to ground of this effective circuit is zero:
\begin{equation}
\nu_{0} = \sqrt{\frac{1}{L\sub{eff}C\sub{eff}}+\frac{1}{L\sub{SQUID}C\sub{eff}}} 
\label{equ:antiRes}
\end{equation}

Here, $C\sub{eff}$ and $L\sub{eff}$ are the effective capacitance and inductance of the resonator and $L\sub{SQUID}$ is the tunable inductance of the SQUID, which (in the balanced case) is given by
\begin{equation}
L_{\textrm{SQUID}}=\frac{\Phi_{0}}{2\pi \Icrit\left|\cos\left(\frac{\pi \Phi}{\Phi_{0}}\right)\right|}.
\label{equ:Lsquid}
\end{equation}
In this expression $\Icrit$ is the critical current of the SQUID, $\Phi$ is the external applied flux and $\Phi_{0}$ is the flux quantum.

Fig.~\ref{fig:vnaFull} shows the center frequency of the measured anti-resonance dip as a function of flux and voltage bias applied to the SQUID. It clearly shows that the frequency is only flux-tunable in the region where no photon emission occurs in Fig. 2c of the main text. Outside of this region the frequency becomes constant. This can be understood by considering that the series resistance of the $RC$ circuit tilts the load line of the voltage bias at the junction. For bias voltages such that $\bias<R\Icrit(\Phi)$ the SQUID stays on its current branch undergoing Bloch oscillations. Then it contributes a finite inductance to the resonant structure making it flux-tunable.

Only when the voltage is sufficiently high, or $\Icrit$ sufficiently suppressed by the flux bias, a voltage drops over the SQUID. Then its effective inductance becomes infinite and does not contribute to the frequency of the anti-resonance anymore. The boundary between the two regions is well described by a function $|\cos(\pi \Phi/\Phi_{0})|$ (dotted line in Fig.~\ref{fig:vnaFull}, rescaled to match the apex of the bright region). This line is also drawn on Fig.~\psd\psdflux~ for comparison. It delimits the region without photon emission corroborating the interpretation given in the main text.

\section{Resistance measurement}
\begin{figure}
	\centering
	\includegraphics[width=0.7\linewidth]{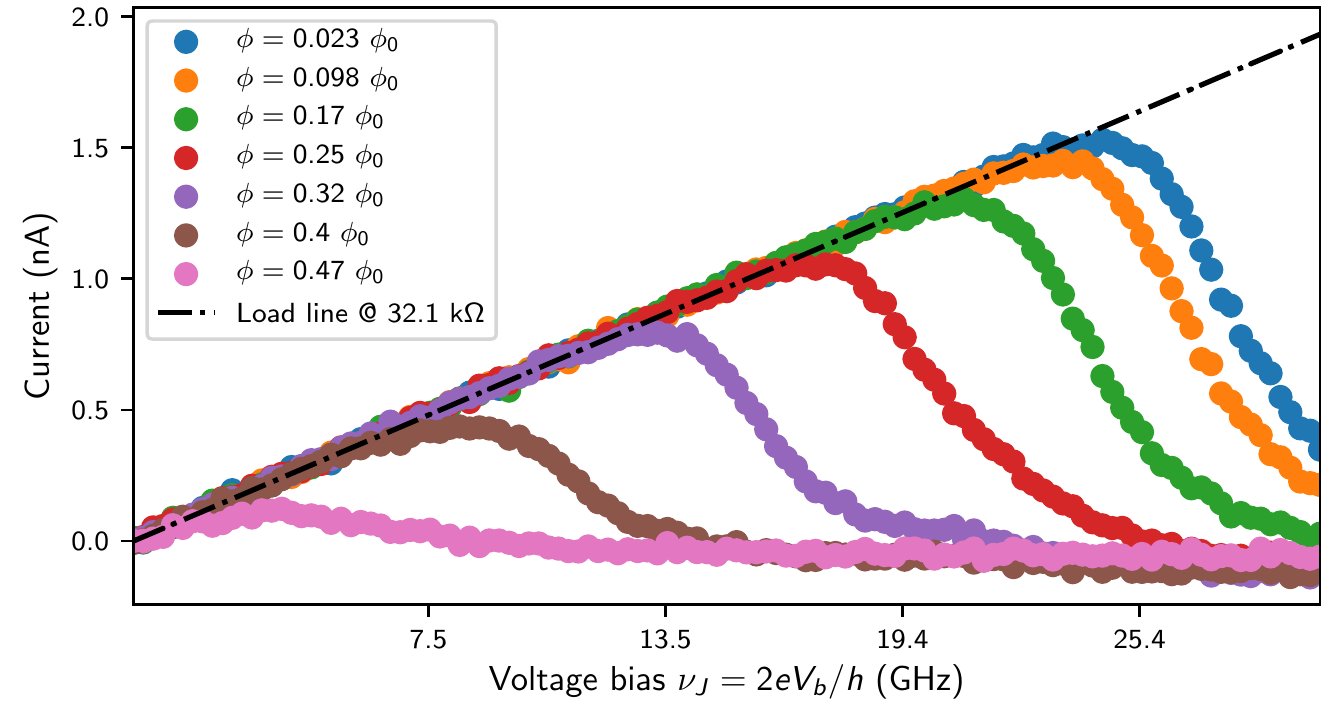}
	\caption{Current through the sample as a function of bias voltage in units of \si{\giga \hertz}. The different curves correspond to different values of flux in the SQUID-loop. The dashed-dotted line shows the fit of the $RC$ circuit resistance.}
	\label{fig:current}
\end{figure}			
The current through the sample as a function of bias voltage was independently measured (see Sec.~\ref{sec:msmt}). The result is given in Fig.~\ref{fig:current}. Each curve was taken at a different flux bias, going from nearly full frustration (lowest current) to minimum frustration (largest current). The slope of the current branch directly gives the resistance in series with the SQUID, which is entirely dominated by the resistor of the $RC$ circuit. From these curves we extract a value of \RExtracted. Moreover, the measured curves show good agreement with the data presented in Sec.~\ref{sss:darkregion}.

\section{Extraction of system parameters using P(E)-theory}
The rate at which Cooper pairs inelastically tunnel through a voltage
biased Josephson junction (or equivalently a voltage biased SQUID acting
as an effective junction with an adjustable critical current) can be
computed with Fermi's Golden Rule considering the junction as a
perturbation to the modes of its electromagnetic environment. The result
depends on correlations between the phase fluctuations at the junction,
which are related to the real part of the environmental impedance seen
by the junction via the fluctuation-dissipation theorem. An overview of
this calculation (called $P(E)$-theory) is given in
reference~\onlinecite{Ingold1992}.

The expression for the Cooper pair tunneling rate into the direction of
the voltage bias is given in Eq.~2 of the main text. It depends on the
function $\PE(h\Jfreq)$ giving the probability density for the
environment to absorb an energy $h\Jfreq$ from a Cooper pair tunneling
through the junction. Here, $\Jfreq=2e\bias/h$ is the energy given to
the  Cooper pair by the voltage bias expressed in frequency units. To
simplify the notation we rewrite $\PE(\Jfreq)=h\PE(h\Jfreq)$. This
function obeys the following normalization and detailed balance relations:

\begin{equation}
1 = \int_{\mathds{R}}\!\mathrm{d}\Jfreq\,\PE(\Jfreq) \approx 
\int_{V}\!\mathrm{d}\Jfreq\,\PE(\Jfreq)
\label{equ:PEnorm}
\end{equation}

\begin{equation}
\PE(-\Jfreq)=e^{-\beta h \Jfreq}\PE (\Jfreq)
\label{equ:PEbalance}
\end{equation}

Here, $\beta = 1/(k\sub{B}T)$, with $k\sub{B}$ the Boltzmann constant
and $T$ the temperature of the electromagnetic environment. The bias 
range $V$ from $\Jfreq = 0$ to \SI{30}{\giga\hertz} of our measurements 
covers the dominant processes at $\Jfreq=E\sub{C}/h\approx\SI{1.5}{\giga 
	\hertz}$, $E\sub{C}/h + f_0 \approx\SI{7.5}{\giga \hertz}$ and 
$E\sub{C}/h + f_1 \approx\SI{19.5}{\giga \hertz}$, so that we can 
consider $P(\Jfreq)$ normalized over the bias range $V$ with good 
approximation.

The second equation signifies that $\PE(-\Jfreq)$ vanishes at zero 
temperature.
This happens, because at negative energies the Cooper pairs have to
tunnel against the voltage bias, thus drawing their energy from the
thermal excitations of the electromagnetic environment. At finite
temperatures $\PE(\Jfreq)$ is given by the integral
equation~\cite{Ingold1992,Minnhagen1976,Grimm2015}:

\begin{equation}
\Jfreq \PE(\Jfreq)=
\frac{2}{R\sub{Q}}\int_{\mathbb{R}}\!\mathrm{d}\freq\,
\PE(\Jfreq-\freq)\frac{\imp }{1-e^{-\beta h\freq}}
\label{equ:Minnhagen}
\end{equation}

This form, called the Minnhagen equation, depends only on the real part
of the frequency dependent environmental impedance \imp, the
temperature, and the superconducting resistance quantum
$R\sub{Q}=h/(4e^{2})\approx\SI{6.5}{\kilo \ohm}$.

It can be shown~\cite{Hofheinz2011, Grimm2015} that the rate density of
photon emission into the electromagnetic environment at frequency
$\freq$ due to forward tunneling Cooper pairs can be expressed in terms
of the critical current of the junction $\Icrit$, the impedance \imp~and
the P(E)-function  as:

\begin{equation}
\gamma(\freq,\Jfreq)=\frac{\Icrit^2}{2h\freq}\imp\PE(\Jfreq-\freq).
\label{equ:PSDsi}
\end{equation}
This quantity is proportional to the power spectral density emitted by
the sample with a conversion factor given by the photon energy $h\freq$
and can be directly measured (see Fig.~\psd\psdfreq~of the main text).
By integrating both sides of the above equation over the voltage bias
and using the normalization property from \eqcite{equ:PEnorm} we find:

\begin{equation}
\int_{V}\!\mathrm{d}\Jfreq'\,\gamma(\freq,\Jfreq')=\frac{\Icrit^2}{2h\freq}\imp.
\label{equ:voltageint}
\end{equation}
The integral is performed over the data shown in Fig.~\psd\psdfreq~of
the main text.
Evaluating \eqcite{equ:PSDsi} at a voltage bias $\Jfreq+\freq$ results in:

\begin{equation}
\gamma(\freq,\Jfreq+\freq)=\frac{\Icrit^2}{2h}\frac{\imp}{\freq}\PE(\Jfreq).
\label{equ:PSDoffest}
\end{equation}

This can be combined with \eqcite{equ:voltageint} to yield the
expression of $\PE(\Jfreq)$

\begin{equation}
\PE(\Jfreq)=\int_\mathbb{R} \!\mathrm{d}\freq\, \sigma_P(f)
\frac{\gamma(\freq,\Jfreq+\freq)}{\int_{V}\!\mathrm{d}\Jfreq'\gamma(\freq,\Jfreq')},
\end{equation}
where $\sigma_P$ is a weight function, which we chose large where
$\gamma$ is large.

\begin{figure}
	\includegraphics[width=0.6\textwidth]{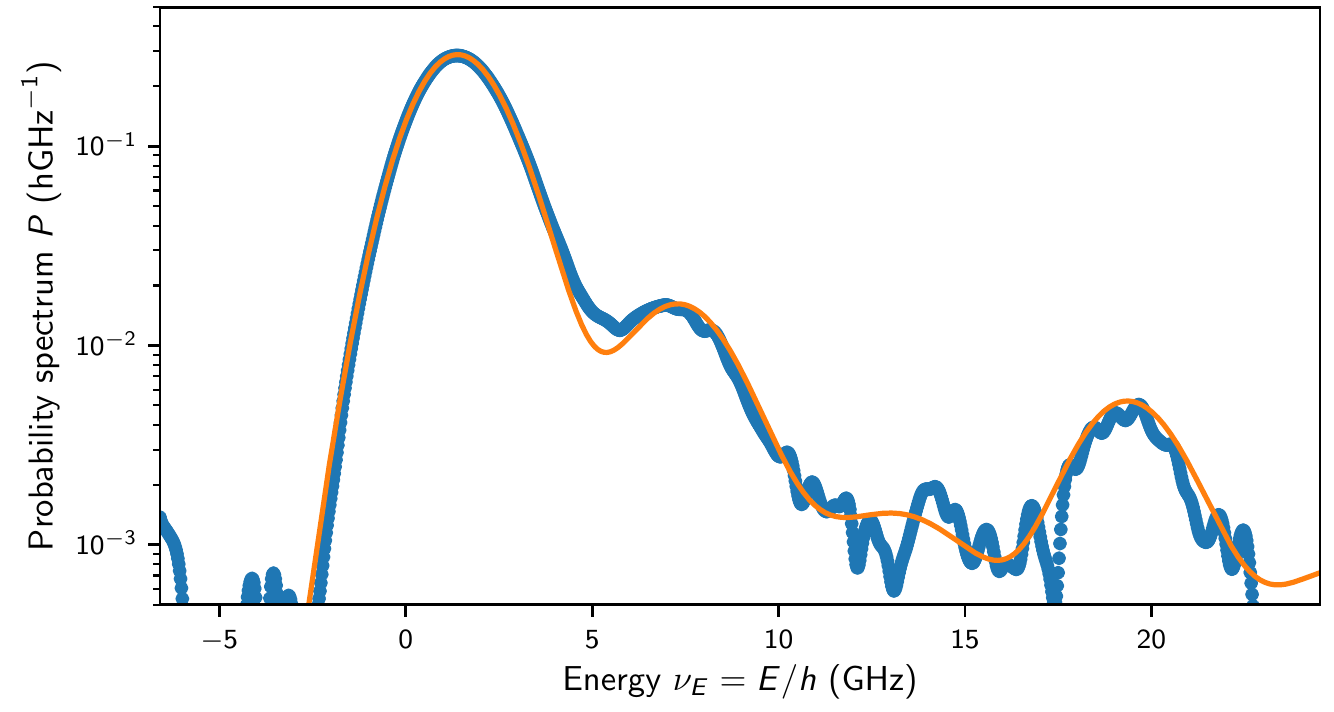}
	\caption{Extracted $\PE(h\Jfreq)$ for forward tunneling Cooper pairs as
		a function of the energy difference a CP has to loose. Negative values
		on the abscissa correspond to thermally activated tunneling against the
		voltage bias. The blue dots are extracted from the data with no
		assumption on the environmental impedance and the orange line is a fit
		of the data based on our impedance model (see Fig.~\overview\circuit).}
	\label{fig:PE}
\end{figure}

Fig.~\ref{fig:PE} shows the extracted $\PE(h\Jfreq)$-function of our
experiment. It displays a prominent peak around
$\Jfreq=E\sub{C}\approx\SI{1.5}{\giga \hertz}$ corresponding to forward
tunneling of Cooper pairs without photon emission while only charging
the capacitance of the $RC$ circuit. This peak does not appear in
Fig.~\psd\psdflux, which only measures the photons emitted around
$\reso$. The next two peaks at \SI{7.5}{\giga\hertz} and
\SI{13.5}{\giga\hertz} are due to the $h\Jfreq=E\sub{C}+h\reso$ and
$h\Jfreq=E\sub{C}+2h\reso$ processes described in the main text. The
second one of these two peaks is linked to a higher order process
(involving two photons at $\reso$) which occurs with much smaller
probability than the first one~\cite{Hofheinz2011}. The last peak at
\SI{19.5}{\giga \hertz} has again a higher amplitude, but does not show
up in Fig.~\psd\psdflux. It comes from the Cooper pair current linked to
the resonance condition $h\Jfreq=E\sub{C}+h\freq\sub{1}$. The photons
emitted by this process are outside of the measurement bandwidth of our
setup.

Note that the photon emission rate density measured at a frequency
$\freq$ and voltage bias $\Jfreq$ such that $\Jfreq+\freq < \freq$,
gives us access to $\PE(\Jfreq)$ for negative arguments. Since this is
related to thermally induced tunneling against the voltage bias, we can
use \eqcite{equ:PEbalance}  to compute $\beta$ and thus the effective
temperature of the electromagnetic environment:

\begin{equation}
\beta = \int_\mathbb{R} \!\mathrm{d} \Jfreq \sigma_\beta(\Jfreq)
\frac{\ln \left( \PE (\Jfreq)/\PE(-\Jfreq)\right)}{h\Jfreq},
\end{equation}
where the weight function $\sigma_\beta$ is chosen largest where we get
the best signal to noise ratio for $\beta$. This temperature together with the
$\PE(\Jfreq)$ function and the measured photon rate density enables us
to extract the critical current. Integrating the Minnhagen equation
(\eqcite{equ:Minnhagen}) and using the normalization property
\eqcite{equ:PEnorm} we get

\begin{equation}
\int_\mathbb{R}\!\mathrm{d}\Jfreq\,\sigma_V(\Jfreq)
\PE(\Jfreq)\Jfreq=\frac{2}{R\sub{Q}}\int_\mathbb{R}\!\mathrm{d}\freq\,
\sigma_f(f) \frac{\imp}{1-e^{-\beta h\freq}}
\label{equ:Minnhint}
\end{equation}
with
\begin{equation}
\sigma_f(\freq) = \int_\mathbb{R}\!\mathrm{d}\Jfreq\,\PE(\Jfreq -
\freq) \sigma_V(\Jfreq).
\end{equation}

\eqcite{equ:voltageint} can be regrouped and integrated over frequency
to give

\begin{equation}
\int_\mathbb{R}\!\mathrm{d}\freq\, \sigma_f(\freq)
\int_{V}\!\mathrm{d}\Jfreq\,
\frac{\gamma(\freq,\Jfreq)\freq}{1-e^{-\beta h\freq}}=
\frac{\Icrit^{2}}{2h}\int_\mathbb{R}\!\mathrm{d}\freq\, \sigma_f(\freq)
\frac{\imp}{1-e^{-\beta h\freq}},
\label{equ:Minnhint2}
\end{equation}

The weight function $\sigma_f$ must be entirely contained in the
measurable bandwidth in order to be able to evaluate these integrals
from our measurement results. In our case this bandwidth is
\SI{4}{\giga\hertz} and is not large enough to fully contain the main
peak of the $\PE(\Jfreq)$ function. We therefore chose $\sigma_V$ to
contain positive and negative values, so that $\sigma_f$, the
convolution product of $\sigma_V$ and $\PE$, remains approximately
limited to our measurement bandwidth.
Through identification of these two equations we then obtain an
expression for the critical current depending only on known quantities
and fundamental constants:

\begin{equation}
\Icrit=4e\sqrt{\frac{\int_\mathbb{R} \!\mathrm{d}\freq\,
		\sigma_f(\freq)\int_{V}\!\mathrm{d}\Jfreq\,
		\frac{\gamma(\freq,\Jfreq)\freq}{1-e^{-\beta
				h\freq}}}{\int_\mathbb{R}\!\mathrm{d}\Jfreq\, \sigma_V(\Jfreq)
		\PE(\Jfreq)\Jfreq}}.
\label{equ:Is}
\end{equation}

Finally, we can extract the impedance seen by the voltage biased SQUID
(shown in Fig.~\psd\psdflux~ of the main text) using
\eqcite{equ:voltageint}:

\begin{equation}
\imp=\frac{2h\freq}{\Icrit^2}\int_{V}\!\mathrm{d}\Jfreq\,\gamma(\freq,\Jfreq).
\label{equ:imp}
\end{equation}

From the photon rate density shown in Fig.~\psd\psdfreq~of the main
article we find an effective temperature of \TeffExtracted~and a
critical current of \IsExtracted.

To verify our approach we now introduce a realistic circuit model based
on the sketch given in Fig.~1a of the main text. It consists of the $RC$
element with resistance $R=\RExtracted$ (measured independently) and
capacitance $C$ (with a parasitic inductance $L\sub{p}$ in series with
the capacitance; see Fig~\overview) as well as a stepped transmission
line resonator with characteristic impedances $Z\sub{0}$ and $Z\sub{1}$.

We perform a fit of the photon rate density using this impedance model
with $C$, $L\sub{p}$, $Z\sub{0}$, $Z\sub{1}$, $T$ and $\Icrit$ as free
parameters. We find the values $Z\sub{0}=\Zzerofit$,
$Z\sub{1}=\Zonefit$, $C=\CExtracted$, $L\sub{p}=\LpExtracted$,
$\Icrit=\IsFit$ and $T=\TeffFit$ in good agreement with our earlier
analysis. The impedance $Z\sub{0}$ of the transmission line section
close to the SQUID is lower than its design value, which could be
explained by the influence of the SQUID capacitance. The shape of $\imp$
given by our circuit model for these values is plotted in
Fig.~\psd\psdflux~ of the main text and reproduces the extracted curve
up to impedance modulations likely due to spurious reflections in our
output lines, which are not part of our model. The
$\PE(h\Jfreq)$-function found by this fit is plotted in Fig.~\ref{fig:PE}.

\end{document}